\documentclass[acmsmall,numbers,prologue,table]{acmart}

\usepackage{amsmath,amssymb,amsfonts}
\usepackage{algorithmic}
\usepackage{graphicx}
\usepackage{textcomp}
\usepackage{xcolor}

\usepackage{enumitem}

\def\BibTeX{{\rm B\kern-.05em{\sc i\kern-.025em b}\kern-.08em
    T\kern-.1667em\lower.7ex\hbox{E}\kern-.125emX}}

\usepackage{xspace}
\newcommand{\etal}{{\em et al.}\xspace}
\newcommand{\eg}{{\em e.g.},\xspace}

\usepackage{url} 
\usepackage{makecell}

\newcommand{\BfPara}[1]{\vspace{1mm}{\noindent\bf#1.}\xspace}
\usepackage{multirow}

\usepackage{pifont}

\usepackage[most]{tcolorbox}

\newcommand{\takeaway}[1]{\begin{tcolorbox}[
  colback=blue!15!white,
  colbacklower=blue!2!white,
  colframe=blue!70!white,
  title=\textbf{Takeaway},
  fonttitle=\sffamily\bfseries,
  enhanced,
  sharp corners=south,
  boxrule=1pt,  
  top=2pt,
  bottom=2pt,
  left=4pt,
  right=4pt,
  enhanced
]
#1
\end{tcolorbox}}

\newcolumntype{L}[1]{>{\raggedright\arraybackslash}m{#1}}
\newcolumntype{C}[1]{>{\centering\arraybackslash}m{#1}}
\newcolumntype{M}[1]{>{\centering\arraybackslash}m{#1}}

\usepackage{array}

\usepackage{xcolor}
\newlength\MAX  
\newcommand*\Chart[1]{#1~\rlap{\textcolor{black!20}{\rule{\MAX}{2ex}}}\rule{#1\MAX}{2ex}}

\usepackage{pgfplots}
\pgfplotsset{compat=1.18}
\usepackage{amsmath}
\AtBeginDocument{%
  \providecommand\BibTeX{{%
    Bib\TeX}}}
    
\usepackage{mdframed}

\definecolor{boxcolor}{RGB}{204, 226, 255} 
\newmdenv[
  backgroundcolor=boxcolor,
  linewidth=0pt,
  innerleftmargin=10pt,
  innerrightmargin=10pt,
  innertopmargin=10pt,
  innerbottommargin=10pt
]{takeawaybox}

\makeatletter
\newcommand{\linebreakand}{%
  \end{@IEEEauthorhalign}
  \hfill\mbox{}\par
  \mbox{}\hfill\begin{@IEEEauthorhalign}
}

\begin{document}

\title{Fishing for Phishers: Learning-Based Phishing Detection in Ethereum Transactions\\}

\author{Ahod Alghuried}
\email{ah104940@ucf.edu}
\affiliation{\institution{University of Central Florida}\city{Florida}\country{USA}}

\author{Abdulaziz Alghamdi}\email{abdulaziz.alghamdi@ucf.edu}
\affiliation{\institution{University of Central Florida}\city{Florida}\country{USA}}

\author{Ali Alkinoon}\email{alialkinoon@ucf.edu}
\affiliation{\institution{University of Central Florida}\city{Florida}\country{USA}}

\author{Soohyeon Choi}\email{soohyeon.choi@ucf.edu}
\affiliation{\institution{University of Central Florida}\city{Florida}\country{USA}}

\author{Manar Mohaisen}
\affiliation{%
  \institution{ Northeastern Illinois University}
  \city{Chicago}
  \state{Illinois}
  \country{USA}
  }
  
\author{David Mohaisen}\email{mohaisen@ucf.edu}\authornote{Corresponding author: David Mohaisen}
\affiliation{\institution{University of Central Florida}\city{Florida}\country{USA}}

\begin{abstract}
Phishing detection on Ethereum has increasingly leveraged advanced machine learning techniques to identify fraudulent transactions. However, limited attention has been given to understanding the effectiveness of feature selection strategies and the role of graph-based models in enhancing detection accuracy. In this paper, we systematically examine these issues by analyzing and contrasting explicit transactional features and implicit graph-based features, both experimentally and analytically. We explore how different feature sets impact the performance of phishing detection models, particularly in the context of Ethereum's transactional network. Additionally, we address key challenges such as class imbalance and dataset composition and their influence on the robustness and precision of detection methods. Our findings demonstrate the advantages and limitations of each feature type, while also providing a clearer understanding of how feature affect model resilience and generalization in adversarial environments.
\end{abstract}

\keywords{Ethereum, Phishing Detection, Transaction Analysis, Machine learning}

 \maketitle
 
\section{Introduction}

Blockchain technology, particularly Ethereum, has revolutionized decentralized transactions, offering secure, transparent, and immutable transaction records~\cite{HeCCHHWCWZ23,LiGL0LXX23, PinioBL24}. However, as blockchain adoption increases, so does its appeal to cybercriminals, with phishing scams emerging as one of the most prevalent forms of attack~\cite{Bibi19,CerneraMMS23, GallettaP24,HeoWYKS23,LiWW0X23, LiGL0LXX23,LinXHZLL23,SaadSNKSNM20, VasekM15}. Phishing scams, which exploit the trust inherent in blockchain transactions and their associated security challenges, account for millions of dollars in losses annually~\cite{ChenGCZL20,HornufMNY2023,SaadCNKM20}. The highly publicized phishing attack on Uniswap Labs in 2022, where attackers stole over eight million dollars, serves as a stark reminder of the risks that Ethereum users face~\cite{YaoZHCVAS2024,SaadCM21,SaadAM19}. Furthermore, recent findings indicate that illicit activities, including phishing, have stolen over two billion USD from Ethereum users, highlighting the urgent need for effective detection mechanisms~\cite{Bibi19,LinXHZLL23,AdeniranHM24}. 

Phishing attacks on Ethereum are distinct from traditional phishing attacks, which typically involve fraudulent websites or emails aimed at stealing sensitive information such as passwords~\cite{YuanHZWZZ20,LiuCWWFZ24,WanXZ23}. In contrast, blockchain phishing often exploits the transparency and pseudonymity of blockchain networks, targeting financial assets directly. These attacks are executed through compromised private keys, deceptive wallet addresses, and malicious smart contracts, enabling unauthorized transactions and asset transfers~\cite{ChenPLLXZ21, KablaAMK22, Zhou2023}. Moreover, while traditional phishing relies on social engineering to deceive users into providing personal information, Ethereum phishing can be automated using scripts that manipulate smart contracts or intercept transactions without direct interaction with the victim~\cite{JonesATN21,BartolettiLLPS21}. This methodological shift underscores the automated nature of threats within the blockchain, necessitating advanced countermeasures~\cite{ChenPNX20,SaadNKKNM19}.

Phishing on blockchain networks such as Ethereum exploits the unique complexities of these systems, using techniques that blur the line between legitimate and malicious transactions~\cite{AlghuriedM24,SaadM23,Zhou2023,SaadKNM21}. The ability of phishing to undermine confidence in blockchain threatens the very foundation of trust and security that these technologies promise. Addressing these risks requires approaches that go beyond traditional methods, incorporating both transactional analysis and advanced machine learning techniques to detect such attacks~\cite{LiGLHLX22, WangCXWSXY22,XuZVQWFLTC23,SaadNKM19}.

\begin{figure}[t] 
\centering 
\includegraphics[width=0.35\textwidth]{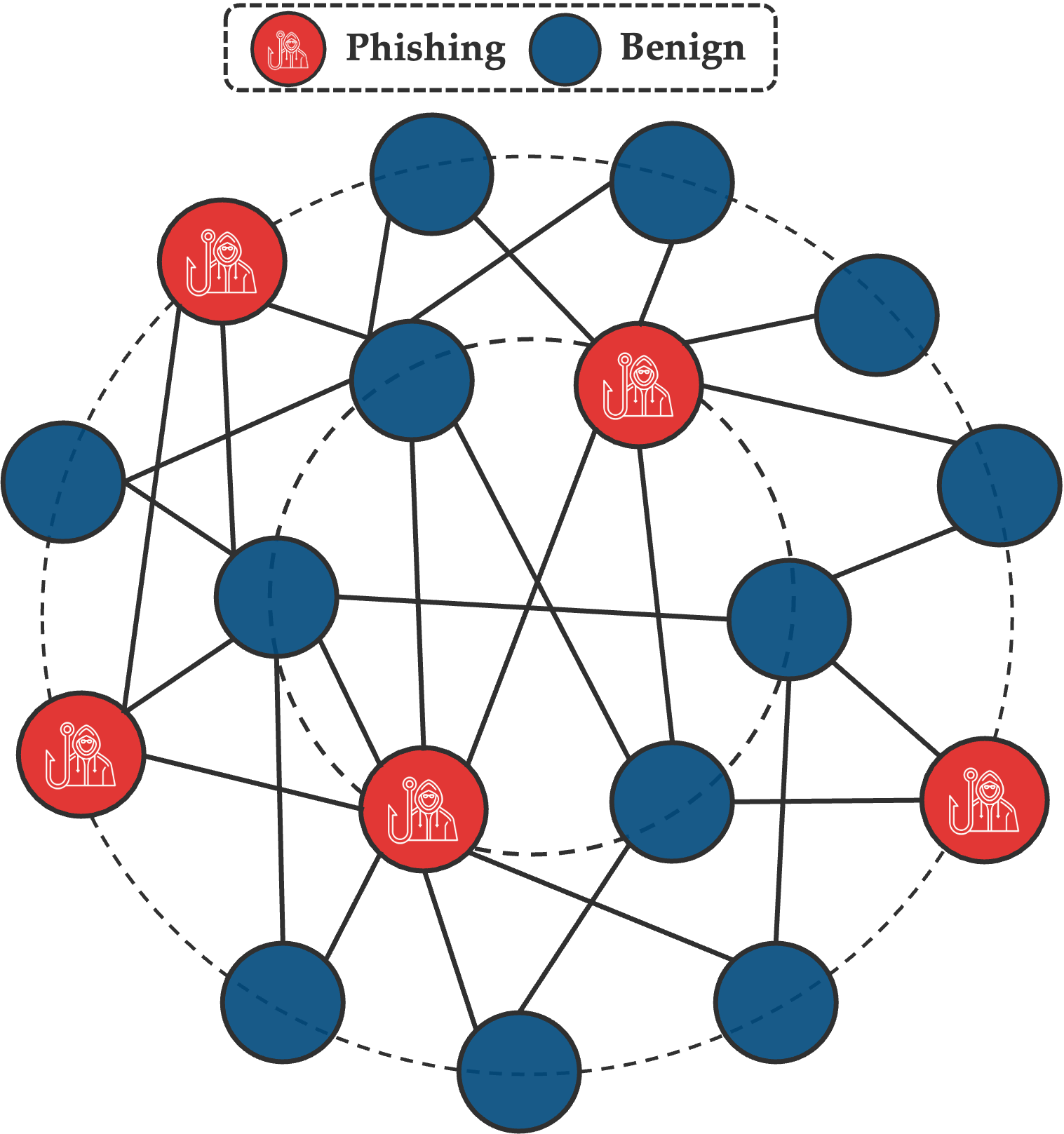}
\caption{Illustration of the Ethereum phishing scam network. Phishing addresses (red nodes) are interspersed among benign addresses (blue nodes), exhibiting similar transactional patterns, thereby complicating detection within the broader network structure.}\label{fig:Ph-B.png}
\end{figure}

Recent advances in phishing detection have increasingly leveraged machine learning models, particularly in blockchain transaction analysis. These models typically rely on explicit transactional features such as transaction values, gas usage, and timestamps~\cite{ChenPLLXZ21, KablaAMK22,SaadAAAYM22}. However, while these features offer valuable insights into individual transaction behaviors, they often fail to capture the broader relational and temporal dynamics essential for detecting phishing~\cite{LiGL0LXX23, AhmadSBM18}. Graph-based approaches, which effectively model the Ethereum transaction network as a graph, have emerged as promising direction for phishing detection. These methods focus on implicit features that reveal interactions between addresses, thus enabling the detection of coordinated and rapid transaction patterns indicative of phishing activity~\cite{WuYLYCCZ22} (see~\autoref{fig:Ph-B.png}). By utilizing advanced analytical methods, researchers aim to develop algorithms capable of discerning subtle signs of illicit activities amid legitimate transactions, addressing both data volume and the cunning nature of these frauds.

Despite progress in this field, current research often overlooks critical issues related to the robustness and scalability of phishing detection models. For instance, many studies focus on achieving high accuracy without addressing the inherent class imbalance in phishing datasets, where phishing transactions are significantly underrepresented compared to benign ones~\cite{FuYF22,Zhou2023,ChenPLLXZ21}. Moreover, there is limited exploration of how feature selection impacts the model’s ability to generalize across diverse phishing scenarios. These challenges underscore the need for more comprehensive and systematic approaches to phishing detection in Ethereum networks.



\BfPara{Contributions} We address these limitations by developing a phishing detection model that systematically and independently evaluates two distinct feature sets: explicit transactional features and implicit graph-derived behavioral features. In contrast to prior work that often aggregates features without assessing their standalone contributions, we adopt a rigorous comparative methodology to isolate and quantify the impact of each feature type on model performance. Our contributions are as follows: 1)  We design and implement a two-stage evaluation framework that contrasts the predictive capabilities of explicit and implicit features in phishing detection, providing insights into their capabilities and limitations. 2) We design a small, focused set of implicit features that describes how Ethereum addresses behave over time, for example, when they send transactions, how often, and on which days. These features go beyond raw transaction values by capturing behavioral patterns that are harder for attackers to fake or hide. 3) We build a Graph Convolutional Network (GCN) that learns from the connections between Ethereum addresses by analyzing how they interact over time and across the network. This allows the model to detect suspicious behavior based on both who is connected to whom and how they behave. 4) We evaluate our model on a large dataset of Ethereum transactions, demonstrating significant improvements in phishing detection performance. 5) Our findings show that a small number of carefully selected implicit features can outperform larger sets of basic transactional features used in prior studies, highlighting that effective feature design is more important than quantity, especially in adversarial settings.

\BfPara{Organization} The remainder of this paper is organized as follows. In \autoref{sec:RW}, we discuss related work on phishing detection. In \autoref{sec:methodology}, we present our data collection process and feature extraction techniques. Our model architecture and experimental setup are outlined in \autoref{sec:experment}. We present our results in \autoref{sec:results}, discussion in \autoref{sec:discussion}, and offer concluding in \autoref{sec:conclusion}.

\section{Related Work}\label{sec:RW}
Phishing detection in Ethereum and other blockchains has attracted considerable research attention as these decentralized systems increasingly become targets for cybercriminals. Various methods have been proposed using explicit transactional and implicit graph-based features. This section reviews the key contributions in the field, organized by feature type, approach, and their relative effectiveness, as summarized in~\autoref{tab:RW}.

\BfPara{Explicit Transactional Features}
Early approaches to phishing detection primarily leverage explicit transactional features extracted directly from blockchain data, such as the number of transactions, gas consumption, timestamps, and transaction values. These features are critical as they provide the initial set of data points that models use to identify potential threats. While these models perform well in detecting clear malicious behavior patterns, they often struggle to capture phishing scams' more complex relational dynamics.

Wen~\etal~\cite{WenXWW23} applied neural networks to explicit transactional features to identify temporal patterns indicative of phishing scams. Similarly, Kabla~\etal~\cite{KablaAMK22} employed features such as from, input, blockHeight, and timeStamp to differentiate phishing accounts from legitimate ones. Despite their success, these models face challenges when dealing with phishing activities that involve sophisticated interactions between multiple blockchain addresses.

Lin~\etal~\cite{LinXHZLL23} refined phishing detection by incorporating explicit transactional features, although their method still encounters difficulties in identifying complex attack structures. Wang~\etal~\cite{WangTPWH24} focused on ransomware detection in the Bitcoin network using features such as the total number of transactions, value exchanged, and the number of neighboring addresses. However, such an approach may fail to detect more elusive behaviors, especially when attackers obscure their operations using multiple addresses.

Yazdinejad~\etal~\cite{YazdinejadDPHKS22} combined explicit transactional data with device activity features in decentralized environments to detect cyber threats. Similarly, Kampers~\etal~\cite{KampersQMV22} used features like trading volume, price fluctuations, and transaction frequency to uncover market manipulation tactics such as spoofing and wash trading. However, reliance on explicit features limits the detection of subtler, network-based strategies.

Ngo~\etal~\cite{NgoWKPAL19} explored the integration of Generative Adversarial Networks (GANs) and anomaly detection to analyze explicit features such as transaction values for phishing detection. While effective in high-dimensional datasets, this method fails to address multi-node phishing attacks. Cheng~\etal~\cite{Cheng0WLL23} introduced a hybrid model that integrates explicit transactional features with a long short-term memory (LSTM) module to capture evolving asset transfer paths, supported by a graph convolutional network (GCN) to analyze their structural properties.

\begin{table}[t]
\centering
\caption{A summary of recent phishing detection studies (since 2021), focusing on their employed features (explicit or implicit), methodological approaches (graph-based or machine learning-based), evaluation metrics (accuracy, precision, recall, and F1-score), and the sizes of the datasets used.}\label{tab:RW}\vspace{-3mm}
\scalebox{0.8}{
\begin{tabular}{lcccllllcc}
\Xhline{2\arrayrulewidth}
\multirow{2}{*}{{Papers}} & \multirow{2}{*}{{Year}} & \multirow{2}{*}{{Features}} & \multirow{2}{*}{{Method}} & \multicolumn{4}{c}{{Performance}} & \multicolumn{2}{c}{{Number of Instances}} \\ \cline{5-10}

  & & & & {Acc.} & {F1-Score} & {Prec.} & {Rec.} & { phishing} & { benign} \\ 
\Xhline{1\arrayrulewidth}

Chen\etal~\cite{ChenPLLXZ21}  & 2021 & Implicit  & Graph-based  & \Chart{0.57} & \Chart{0.23} & \Chart{0.72} & \Chart{0.14} & 1,157 & 2,973,382 \\ \hline
Xia~\etal~\cite{XiaYLJW22}    & 2022 & Implicit  & Graph-based  & -     & \Chart{0.81}   & \Chart{0.81}   & \Chart{0.82}   & 451   & 12,834   \\ \hline
Kabla~\etal~\cite{KablaAMK22}   & 2022 & Explicit  & ML-based     & \Chart{0.98}   & \Chart{0.98}   & \Chart{0.98}   & \Chart{0.98}   & 5,448 & 79,216   \\ \hline
Li~\etal~\cite{LiGLHLX22}    & 2022 & Implicit  & ML-based     & \Chart{0.92} & \Chart{0.81}  & \Chart{0.77}  & \Chart{0.85}  & 4,932 & 6,844,050 \\ \hline
Wu~\etal~\cite{WuYLYCCZ22}   & 2022 & Implicit  & ML-based     & -      & \Chart{0.90}    & \Chart{0.92}  & \Chart{0.89}   & 1,259 & 1,259    \\ \hline
Fu~\etal~\cite{FuYF22}       & 2022 & Implicit  & Graph-based  & \Chart{0.88}   & \Chart{0.87}   & -      & -    & 1,928 & 1,901    \\ \hline
Zhou~\etal~\cite{Zhou2023}     & 2023 & Explicit  & Graph-based  & \Chart{0.98}  & \Chart{0.97}   & \Chart{0.96}  & \Chart{0.99}  & 1,659 & 5,805    \\ \hline
Lin~\etal~\cite{LinXHZLL23}   & 2023 & Explicit  & ML-based     & \Chart{0.82}   & \Chart{0.82}   & \Chart{0.87}  & -     & 301   & 4,116,315 \\ \hline
Li~\etal~\cite{LiGL0LXX23}   & 2023 & Implicit  & Graph-based  & -    & \Chart{0.92}   & \Chart{0.91}   & \Chart{0.92}  & 5,639 & 25,000   \\ \hline
Cheng~\etal~\cite{ChengZWLL24}  & 2024 & -         & ML-based     & \Chart{0.96}  & \Chart{0.68}  & \Chart{0.62}  & \Chart{0.75}  & 7,696 & 89,318   \\ \hline
Liu~\etal~\cite{LiuCWWFZ24}   & 2024 & Implicit  & Graph-based  & -      & -      & -      & -      & 5,363 & 330,000  \\ \hline
Our work  & 2024 &  Implicit & Graph-based  &  \Chart{0.95}     &  \Chart{0.95}     &  \Chart{0.96}    & \Chart{0.95}    & 671,865 & 2,687,460   \\ \hline
\end{tabular}}
\end{table}

\BfPara{Implicit Graph-Based Features}
In contrast to explicit feature-based methods, implicit approaches focus on capturing the relational dynamics inherent in blockchain networks. These methods frequently employ Graph Neural Networks (GNNs) or similar graph-based models to analyze the network structure between addresses, enabling the detection of more phishing schemes.

Zhou~\etal~\cite{Zhou2023} introduced an approach called the Edge-Featured Graph Attention Network (EGAT), which leverages both node and edge features, such as transaction values, gas usage, and timestamps, to detect phishing behavior by focusing on the relationships between network nodes. This method uncovers hidden patterns that transactional feature-based models often miss, offering a deeper insight into the complex interplay of network interactions.


Li~\etal~\cite{LiGL0LXX23} proposed the Transaction Graph Contrast Network (TGC), which utilizes contrastive learning to improve phishing detection by leveraging robust representations of Ethereum addresses within transaction subgraphs. TGC enhances its detection capabilities by introducing node-level and context-level contrast modules, making it particularly effective in large, dynamic networks.

\BfPara{Limitations and Our Contribution} Despite notable progress, existing approaches still face key limitations. Models that rely only on explicit features~\cite{KablaAMK22,WenXWW23} often miss the broader network context that can reveal important phishing patterns. Conversely, graph-based models~\cite{LiGL0LXX23} may overlook transactional behaviors that are essential for understanding user activity. Our work addresses both issues by first evaluating explicit features to capture direct behavioral patterns, and then analyzing implicit, graph-based features to understand the relational structure between addresses. Unlike prior studies, we evaluate the model in two distinct phases—using explicit and implicit features separately—to better assess their individual impact on performance. We also introduce fine-grained temporal features, such as the time difference between consecutive transactions, which highlight behavioral anomalies that are often missed in existing graph-based approaches. To validate the importance of these features, we use a Random Forest classifier (RF) to assess their predictive value in distinguishing phishing from benign activity. In addition, our study highlights the most informative implicit features. It incorporates a weighted loss function to address the class imbalance, resulting in improved detection performance across multiple metrics.

\begin{figure}[t] 
\centering 
\includegraphics[width=0.70\textwidth]{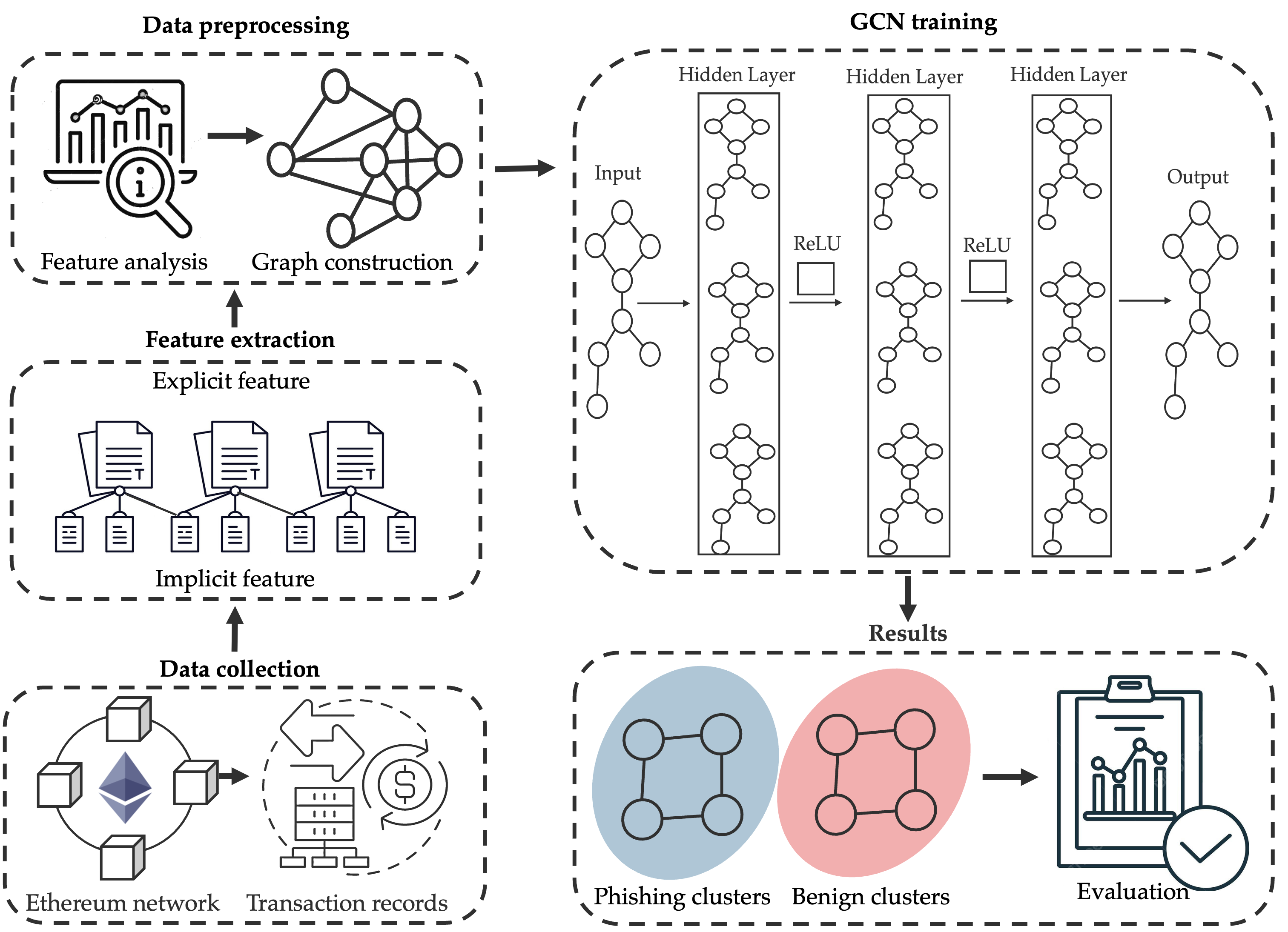}
\caption{An illustration of the proposed pipeline, integrating explicit and implicit features from the Ethereum network.} 

\label{fig:pipline}
\end{figure}

\section{Methodology}\label{sec:methodology}


Our pipeline, shown in~\autoref{fig:pipline}, involves several steps as follows. First, Ethereum transactions are collected and labeled as phishing or benign. Then, both explicit and implicit features are extracted from the data. Weighted loss functions are applied to address the class imbalance. Subsequently, these features are fed into GCN, aggregating information from neighboring nodes to classify addresses. Finally, the model's performance is evaluated using key metrics. In the following, we elaborate on the various steps in our pipeline.

\subsection{Data Collection}

The primary goal of data collection was to compile a comprehensive dataset of Ethereum transactions, explicitly focusing on phishing-related and benign activities. This dataset forms the foundation for developing and testing models to detect phishing transactions on the Ethereum network. The collected data includes transactions associated with known phishing addresses and benign transactions, providing a balanced view for robust model training.

\BfPara{Data Collection Process} To gather the transaction history for each phishing address, the Etherscan API\footnote{Accessible at: \url{https://etherscan.io} (accessed November 2024).} was employed. The API provided access to detailed transactions, including block numbers, timestamps, sender and receiver addresses, and transaction values, which are essential for analyzing blockchain activities. The phishing addresses were sourced from a publicly available dataset on GitHub\footnote{Accessible at: \url{https://github.com/YNclusk/scamsonethereum} (accessed November 2024).}, compiled initially and utilized in~\cite{Kimber2023}. This dataset includes  7,915 unique Ethereum addresses that have been flagged for their involvement in phishing activities. These addresses were verified using the Etherscan API to confirm their involvement in phishing activities. Each address was cross-referenced against transaction histories and community reports to ensure the accuracy of their classification as malicious. This rigorous process supports our dataset reliability by substantiating the addresses phishing history.

A Python script was designed to automate the data collection process by querying the Etherscan API for transaction data associated with identified phishing addresses. The script systematically extracted details such as block numbers, timestamps, and transaction values, and organized them into a structured CSV file for efficient analysis.

Benign transactions, defined as those not associated with the identified phishing addresses, were carefully collected using the same API parameters and methods. Using the identical API parameters and collection methods, the benign transactions were directly comparable to the phishing transactions regarding data structure and content.

Once the phishing and benign transactions were collected, they were combined into a single dataset. The dataset was labeled to differentiate between phishing and benign transactions. Transactions associated with the 7,915 phishing addresses were labeled as phishing, while those related to other addresses were labeled as benign. The final dataset includes the following features (explained in \autoref{tab:featuers}): Timestamp (G1), Transaction Hash, Sender Address (G2), Receiver Address (G3), Transaction Value (G4), Gas Used (G7), Gas Price (G6) and Label. The dataset's composition concluded with around 2M benign and around 600K phishing transactions, providing a substantial basis for our phishing detection model.


\begin{figure}[t] 
\centering 
\includegraphics[width=0.70\textwidth]{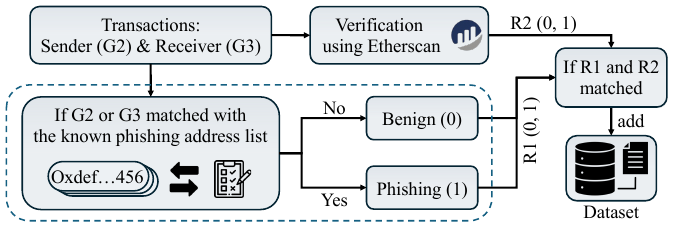}
\caption{Illustrating the labeling process. G2 and G3 represent the sender and receiver addresses. R1 and R2 are labeling results. In R1, addresses are first matched against the known phishing list. If a match is found, R2 performs manual verification via Etherscan to ensure labeling accuracy.}

\label{fig:labeling-process}
\end{figure}

\BfPara{Labeling Strategy} 
Each Ethereum transaction in this study was labeled as either phishing (1) or benign (0). The labeling process is illustrated in ~\autoref{fig:labeling-process}. G2 and G3 represent the sender and receiver addresses, while R1 and R2 are intermediate labeling results. In the first step (R1), an address is matched against the known phishing list compiled from a publicly available dataset used in prior work~\cite{Kimber2023}. If a match is found, a second step (R2) involves manual verification through the Etherscan API to confirm phishing activity and reduce labeling errors.

Only transactions involving addresses directly (1-hop) connected to verified phishing addresses were labeled as phishing. This conservative approach minimizes false positives, though it may overlook some phishing behaviors. For example, if address \texttt{"0xabc...123"}, listed in the phishing dataset, sends funds to \texttt{"0xdef...456"}, that transaction is labeled as phishing.

For benign labels, we used addresses with no known history of phishing, scams, or related activity. Although some benign addresses may later be flagged as malicious, the risk is minimized by selecting historically clean addresses.  For example, if address \texttt{"0x111...aaa"} sends funds to \texttt{"0x222...bbb"}, and neither address appears in the phishing list, the transaction is labeled as benign. To ensure consistency and fairness, both phishing and benign transactions were collected using the same Etherscan API and stored in a unified data format.

\BfPara{Data Balancing} A weighted loss function was employed to address the significant class imbalance in the dataset, where phishing transactions constituted only 7.63\% of the total transactions. This approach involved assigning higher weights to the underrepresented phishing class and lower weights to the more prevalent benign class. This weighting strategy compels the model to focus more on phishing, despite their relative scarcity, thereby improving the model's sensitivity to phishing. The loss function is defined as follows:

\begin{equation}
\mathcal{L} = - \sum_{i=1}^{N} w_{y_i} \cdot \log p(y_i) \label{eq:lossfunction} 
\end{equation}

where \(N\) is the number of nodes, \(y_i \in \{0, 1\}\) is the true class label of node \(i\), \(p(y_i)\) is the model’s predicted probability for the true class, and \(w_{y_i}\) is the weight assigned to the class.

A weighted loss function (\autoref{eq:lossfunction}) was employed to address the problem of class imbalance in the dataset, as phishing nodes occur much less frequently compared to benign nodes. This method increases the penalty for incorrectly classifying phishing nodes, encouraging the model to give special attention to these rare but important cases. As a result, the model avoids overly favoring the majority class and improves its ability to detect phishing behavior effectively. The weighted loss function was selected over alternatives like focal loss because it is simpler to implement, provides better stability during training, and does not require tuning additional parameters. Furthermore, it naturally aligns with graph neural network models, allowing effective learning from imbalanced data without adding unnecessary complexity. This choice is particularly valuable in phishing detection tasks, where overlooking phishing nodes typically causes more harm than mistakenly flagging benign ones. Ultimately, the use of a weighted loss function helps the model to become more sensitive to the minority class, leading to more balanced and accurate results.

\subsection{Feature Selection Rationale}
The selection of features for this study was guided by a combination of observational insights and established research on blockchain behaviors~\cite{ChenPLLXZ21,Zhou2023,LiuCWWFZ24}. This informed approach ensures that the features we focus on are both indicative of phishing activities and reflective of the unique dynamics within blockchain transactions.

\BfPara{Transaction Volume (G4)} Transaction volume is a critical indicator in phishing detection. Phishing transactions often involve volumes that are either significantly higher or lower than those of typical benign transactions. High-value transactions may be executed with the intent to quickly drain compromised wallets, capitalizing on the rapid execution capabilities of blockchain technologies. Conversely, attackers might also distribute funds in numerous smaller transactions to mimic routine user behavior, thereby evading detection systems that are tuned to spot large, irregular transfers. This dichotomy in transaction sizes provides crucial signals for distinguishing between benign and malicious activities within the network.

\BfPara{Gas Usage (G7)} Manipulation of gas usage is a common tactic in phishing schemes, utilized to optimize the execution success of malicious transactions. High gas fees are often prioritized to ensure fraudulent transactions are processed swiftly, outpacing any reactive security measures. On the other end of the spectrum, lower-than-average gas fees can be indicative of an attacker's strategy to minimize operational costs during extensive phishing campaigns that require the execution of numerous transactions. By analyzing patterns in gas usage, our model can identify deviations from the norm that suggest underlying phishing attempts.

\BfPara{Timing Features (G1)} The timing of transactions offers profound insights into user behavior on the blockchain. Phishing operations frequently exhibit abnormal timing patterns—such as sudden bursts of high-intensity activity followed by extended periods of dormancy—that starkly contrast with the more uniform transaction timing of regular users. These anomalies in transaction timing are vital for identifying potential phishing activities, as they often reflect the opportunistic nature of attacks and the subsequent attempts to hide illicit actions within normal traffic flows.

\BfPara{Node Connectivity (Implicit Features)} The relational dynamics between nodes, or addresses, in the blockchain provide a ground for detecting phishing. Patterns such as repetitive transactions with certain clusters of addresses or the sudden emergence of transactions with new, previously unrelated nodes can signal the operation of controlled accounts engaged in phishing. These connectivity patterns, especially when they deviate from typical user behavior, are strong indicators of coordinated malicious activities.

The efficacy of implicit features in our phishing detection framework hinges on their ability to uncover subtle yet consistent anomalies in transaction patterns and inter-node relationships. The theoretical underpinnings and empirical validations from previous studies bolster our reliance on graph-based features to robustly detect anomalies in network security~\cite{ChenPLLXZ21,XiaYLJW22,LiGLHLX22}. Moreover, the inherent transparency of blockchain transactions allows for a comprehensive analysis of these relational patterns, rendering these implicit features especially powerful for uncovering and understanding the sophisticated strategies employed in phishing attacks. This detailed exploration not only aids in effectively identifying phishing activities but also enhances our understanding of the interaction paradigms within Ethereum's complex system.

\subsection{Feature Extraction}
This work employs a GCN to detect phishing addresses on the Ethereum blockchain by testing the model with two feature sets: explicit transactional features and then implicit graph-based features. The objective is to compare which feature set is more effective in improving the model’s performance and delivering better results. Exploring these different aspects allows us to understand the impact of each feature type on the predictive capabilities of the model, providing insights into optimizing detection for enhanced accuracy and efficiency.

\BfPara{Explicit Features} The explicit features refer to the direct, transaction-specific attributes extracted from the raw Ethereum data, providing key insights into the fundamental properties of each transaction. These features include critical fields such as transaction timestamp (G1), sender (G2) and recipient (G3) addresses, value transferred (G4), and gas-related features (G5, G6, G7). The goal of leveraging explicit features is to capture fundamental transactional behavior, such as the timing, volume, and cost-efficiency of transactions, which can provide early indicators of phishing. This direct approach to feature extraction helps in quickly assessing transaction integrity and forming a preliminary defense against phishing tactics.

\BfPara{Implicit Features} To capture the deeper structural and temporal dynamics within the Ethereum transaction, implicit features were extracted from the transaction graph. Unlike explicit features, which focus on individual transaction details, implicit features highlight patterns from the interactions between nodes (addresses) over time, offering a broader perspective on behavior. These features are essential for understanding the complex network relationships and potential collusion among addresses that are often characteristic of phishing.

The implicit features analyze how nodes interact within the network, uncovering patterns like bursts of rapid transactions followed by inactivity, often associated with phishing activities. Time-based features such as the average time between transactions (G1), the standard deviation of transaction times (G1), and identifying irregular transaction patterns common in phishing schemes.

A crucial component of implicit features is the examination of node relationships. For instance, repetitive or coordinated behaviors, such as frequent small-value transfers, can indicate phishing clusters. Considering node behavior within the entire transaction graph allowed for identifying more subtle patterns linked to phishing. Temporal characteristics also proved valuable in distinguishing phishing from benign addresses. Metrics such as transaction time intervals and the duration between a node’s first and last transactions helped detect anomalies in timing. A detailed description of the explicit and implicit features is provided in~\autoref{tab:featuers}. These analyses provide a deeper insight into the tactics employed by attackers, enabling more effective prevention and mitigation strategies.

\begin{table}[t]
\centering
\caption{Summary of explicit and implicit features transactions on the Ethereum network.}
\label{tab:featuers}

\scalebox{0.85}{\begin{tabular}{l|l|l|l}
\hline
{T} & {Features Used} &{G}& {Explanation} \\ \hline
\multirow{7}{*}{\rotatebox[origin=c]{90}{Explicit Features}} & 
{TimeStamp} & G1  & The timestamp of the transaction \\ \cline{2-3}
& {From} & G2 & The sender's address \\ \cline{2-3}
& {To} & G3  & The recipient's address \\ \cline{2-3}
& {Value} & G4  & The amount of Ether transferred \\ \cline{2-3}
& {Gas} & G5  & The amount of gas provided for the transaction \\ \cline{2-3}
& {GasPrice} & G6  & The price of gas (in Wei) used for the transaction \\ \cline{2-3}
& {GasUsed} & G7  & The actual amount of gas used in the transaction 
\\ \hline

\multirow{16}{*}{\rotatebox[origin=c]{90}{Implicit Features}} & 
{From\_tx\_cnt} & G2  & Number of transactions initiated by address \\ \cline{2-3}
& {To\_tx\_cnt} & G3 & Number of transactions received by address \\ \cline{2-3}
& {Total\_val\_sent} & G4 & Total Ether sent by address \\ \cline{2-3}
& {Total\_val\_rec'd} & G4  & Total Ether received by address \\ \cline{2-3}
& {Avg\_gas\_sent} & G7 & Average gas for transactions sent \\ \cline{2-3}
& {Avg\_gas\_rec'd} & G7 & Average gas for transactions received \\ \cline{2-3}
& {Mean\_hour\_sent} & G1  & Average hour of day when transactions are sent \\ \cline{2-3}
& {Mean\_hour\_recd} & G1  & Average hour of day when transactions received \\ \cline{2-3}
& {Std\_hour\_sent} & G1  & Standard deviation of hour transactions sent \\ \cline{2-3}
& {Std\_hour\_recd} & G1  & Standard deviation of hour transactions received \\ \cline{2-3}
& {Avg\_time\_bw\_tx} & G1 & Average time b/w consecutive transactions sent \\ \cline{2-3}
& {Min\_time\_bw\_tx} & G1  & Minimum time b/w consecutive transactions sent \\ \cline{2-3}
& {Max\_time\_bw\_tx} & G1  & Maximum time b/w consecutive transactions sent \\ \cline{2-3}
& {Tx\_duration} & G1  & Duration b/w  first and last transactions of address \\ \cline{2-3}
& {Wd\_tx\_ratio\_sent} & G1 & Proportion of transactions sent on weekends \\ \cline{2-3}
& {Wd\_tx\_ratio\_recd} & G1  & Proportion of transactions received on weekends \\ \hline
\end{tabular}}
\end{table}

\BfPara{Feature Analysis and Selection} To identify the most distinguishing features for classifying phishing and benign nodes, we utilized statistical analysis and Random Forest (RF). {Random Forest is an ensemble learning method that constructs multiple decision trees during training and determines the output class based on the majority vote among the trees. It is widely used due to its robustness, interpretability, and ability to handle high-dimensional data. Statistical analysis was first employed to compare key features between phishing and benign nodes, such as transaction amounts, gas usage, and transaction timing. This helped uncover significant behavioral differences between the two types of nodes. The RF classifier was then applied to rank the importance of the extracted features. This dual approach ensures a comprehensive understanding of feature relevance, enhancing the model’s accuracy by focusing on the most predictive attributes. RF provides a quantifiable measure of feature importance, highlighting the most critical features for distinguishing phishing nodes from benign ones. This method validated the statistical analysis findings and helped refine the feature set by prioritizing attributes that proved to be key indicators of phishing behavior.}


\BfPara{Data Preprocessing} Before applying the model, several preprocessing steps were performed to ensure the data was in an appropriate format and range for effective and robust model training. Namely, these steps included normalization and scaling. These preprocessing efforts align data from diverse sources and scales, enhancing model training and performance predictability.

\noindent {\ding{172} \em Normalization:} To alleviate the bias due to different feature scales, the Min-Max scaling was applied to the features. This scaling ensures that all features are transformed into a [0,1] range, which then aids in improving convergence during training. Normalization is particularly important in handling outliers and reducing skewness in data distribution. The {\tt MinMaxScaler} from {\tt scikit-learn} was utilized to rescale each feature based on its minimum and maximum values within the dataset.

\noindent{\ding{173} \em Scaling:} The features were categorized into explicit (\eg timestamp, value) and implicit features (\eg transaction frequency, behavioral patterns). After scaling, these features were integrated into a node feature matrix that was used as input for the GCN model. This structured approach to feature integration facilitates more effective learning by the neural network, optimizing the detection of complex phishing patterns.


\begin{figure}[t] 
\centering 
\includegraphics[width=0.70\textwidth]{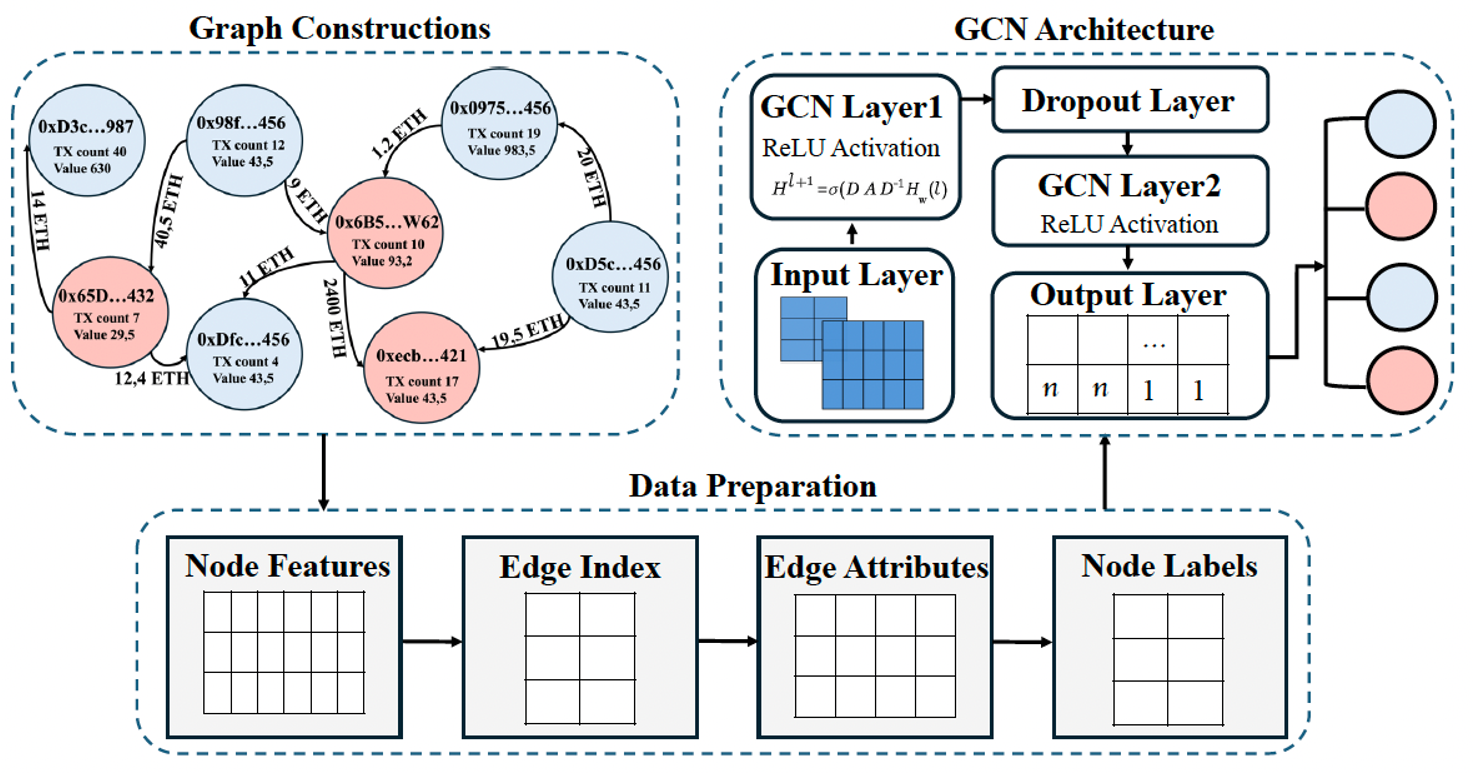}
\caption{{Construction of a directed Ethereum transaction graph and its transformation into PyTorch Geometric inputs. The resulting data is processed by a GCN to classify addresses as phishing or benign.}}

\label{fig:GCN}
\end{figure}

\subsection{Graph Construction}

A directed graph was constructed using {\tt NetworkX}~\cite{networkx}, where each node represents an Ethereum address and each edge represents a transaction between two addresses. {~\autoref{fig:GCN} illustrates the full pipeline, including graph construction, data preparation, and the GCN architecture used for address classification.} This graph captures the underlying structure of the Ethereum transaction network and allows for a detailed analysis of how Ethereum addresses interact. Understanding these interactions is essential for visualizing complex network dynamics and serves as a foundation for analytical models. After construction, the graph is transformed into a format compatible with PyTorch Geometric for training the GCN. The conversion involves:

\begin{enumerate}
    \item {\bf Node Features.} A matrix where each row corresponds to the feature vector of a node, representing the explicit and implicit attributes of the Ethereum addresses. This matrix supplies the necessary data for the GCN to evaluate each node based on its distinct characteristics.

\item {\bf Edge Index.} A tensor representing the directed edges between nodes, indicating the relationships between the sender and the receiver addresses (edge). This tensor is vital for the GCN to recognize and utilize the connections between nodes, facilitating effective feature propagation through the network.
    
\item {\bf Edge Attributes.} A label indicating whether the node is involved in phishing activity (1 for phishing, 0 for benign). These labels are imperative for training the model to accurately classify nodes based on their transactional behaviors and associations.
    
\item {\bf Node Labels.} A label indicating whether the corresponding labeled node is phishing 1 or benign 0. This classification supports the supervised learning process, guiding the GCN in generating precise predictive outcomes.
\end{enumerate}

\BfPara{Model Architecture} A GCN was implemented to classify Ethereum addresses as phishing or benign. The GCN aggregates features from neighboring nodes and propagates information through the graph, enabling the model to learn embeddings for each node by considering both its features and those of its neighbors. This architecture takes advantage of network connectivity and feature sets to uncover subtle indicators of malicious activities that traditional methods might miss.

\begin{enumerate}
\item {\bf Input Layer.} The node feature matrix, where each node is represented by a vector of explicit features. This layer is the entry point for data into the GCN, setting the foundation for complex pattern detection.

\item {\bf Hidden Layers.} Multiple GCN layers that apply graph convolutions to propagate information between neighboring nodes. Each GCN layer updates the representation of a node by aggregating the features of its neighbors. These layers refine the raw data into actionable insights, crucial for the detection process.

\item {\bf Activation Functions.} {\tt ReLU} (Rectified Linear Unit)  activation is applied after each hidden layer to inject non-linearity into the model, enhancing its capability to model complex relationships.

\item {\bf Dropout.} Dropout regularization is strategically applied to the hidden layers to effectively prevent overfitting, especially in the presence of highly imbalanced data. This technique ensures that the model remains generalizable and effective against various forms of data variance.

\item {\bf Output Layer.} The output layer uses a softmax activation function to produce the final classification (phishing or benign) for each node. This layer determines the ultimate classification outcome, translating learned embeddings into definitive labels.
\end{enumerate}

\BfPara{Graph Construction} The Ethereum transaction network was modeled as a directed graph \( G = (V, E) \), where:

\begin{itemize}
    \item[\ding{71}] {\bf Nodes.} \( V \) represents the nodes corresponding to Ethereum addresses. Each node is enriched with various features, either explicit or implicit, capturing important aspects of the transaction behavior. 
    
    \item[\ding{71}] {\bf Edges.} \( E \) represents the directed edges corresponding to transactions between addresses. Each edge connects a sender node \( u \) to a receiver node \( v \), representing a transaction from \( u \) to \( v \), with additional attributes.
\end{itemize}

\BfPara{Graph Convolutional Layers} The core of the GCN model consists of graph convolutional layers, which operate on the adjacency matrix \( A \) of the transaction graph and the feature matrix \( X \), with a feature vector for each node. The GCN aggregates the features of each node’s neighbors, propagating the aggregated information through multiple layers. This enables the model to learn implicit features that capture the relationships between addresses.

The \textbf{layer-wise propagation rule} for the GCN is given by:

\begin{equation}
H^{(l+1)} = \sigma \left( D^{-\frac{1}{2}} A D^{-\frac{1}{2}} H^{(l)} W^{(l)} \right) \label{eq:gcn-propagation}
\end{equation}


Where \( A \) is the adjacency matrix of the graph, representing connections between nodes, \( D \) is the degree matrix, where each diagonal element represents the degree of the corresponding node, \( H^{(l)} \) is the hidden state at layer \( l \), representing the node embeddings at that layer, \( W^{(l)} \) is the weight matrix for layer \( l \), which is learned during training, and \( \sigma \) is an activation function, ReLU in this case, applied element wise.

\BfPara{Evaluation Metrics} The performance of the model was assessed using the following key metrics, each defined mathematically.

\begin{enumerate}
    \item[\ding{172}] {\em Accuracy:} Accuracy measures the overall proportion of correctly classified phishing and benign nodes. It is defined as: $\text{Accuracy} = \frac{TP + TN}{TP + TN + FP + FN}$, where \( TP \) (True Positives) are correctly identified phishing, \( TN \) (True Negatives) are correctly identified benign, \( FP \) (False Positives) are benign misclassified as phishing, and \( FN \) (False Negatives) are phishing misclassified.

    \item[\ding{173}] {\em Precision:} Precision evaluates the accuracy of the phishing node predictions, focusing on minimizing false positives. It is defined as: $\text{Precision} = \frac{TP}{TP + FP}$. Precision reflects the proportion of correctly identified phishing nodes out of all nodes predicted as phishing.

    \item[\ding{174}] {\em Recall:} Also known as sensitivity or true positive rate, recall measures the model's ability to detect actual phishing nodes. It is defined as: $\text{Recall} = \frac{TP}{TP + FN}$. Recall emphasizes the proportion of actual phishing nodes that were correctly identified by the model.
    
    \item[\ding{175}] {\em F1-Score:} The F1-score provides a balanced measure of precision and recall as the harmonic mean of the two and is useful when there is an imbalance between the classes. The F1-score is defined as: $\text{F1-Score} = 2 \times \frac{\text{Precision} \times \text{Recall}}{\text{Precision} + \text{Recall}}$. This metric is valuable in assessing performance when both precision and recall are critical.
\end{enumerate}

{\subsection{System Model}}

{This work presents a behavior-based detection system that operates at the Ethereum address level. Rather than focusing on transactional attributes, the system is built around implicit behavioral characteristics—subtle, often hidden patterns that are difficult for adversaries to forge or predict. The goal is to distinguish phishing addresses from benign ones by modeling how addresses behave over time and within their transactional context.}

{The system starts by constructing a directed transaction graph, where each node represents an Ethereum address, and each edge represents a transaction between two addresses. This graph is the basis for learning behavioral features. Each node is embedded with fine-grained behavioral signals derived from its historical activity, including how frequently it transacts, the timing and variability of its transactions, and its interaction rhythm with other addresses.}

{Key features include average inter-transaction times, weekend activity ratios, and standard deviation in transaction hours—attributes that together form a behavioral fingerprint of each address. These patterns reveal how an address typically operates and whether its behavior aligns more closely with legitimate users or with patterns common to phishing activity. For instance, some phishing addresses stayed inactive for long periods and then started sending a series of low-value transactions during early morning hours (\eg, 2–4 AM UTC). These transactions were usually sent to new or unfamiliar addresses, which may suggest testing or hiding funds. These actions may not seem suspicious. However, when combined with short gaps between transactions, regular timing, or sudden changes in gas prices, the overall pattern can point to automated or hidden activity that is less common among normal user behavior.}

{To capture these relationships, the system applies GCN to model these behavioral relationships by aggregating feature information from neighboring nodes in the transaction graph. This mechanism allows the model to learn from the features of each address in isolation and the transactional context in which it operates, such as repeated interactions with certain nodes, timing irregularities, or sudden shifts in activity patterns. Additionally, the final embedding for each address is passed through a classification layer that outputs a phishing probability score. Addresses exceeding a fixed threshold are classified as phishing, enabling proactive intervention. The system is also designed to operate passively on historical transaction data without relying on future information. Classification decisions are made at the address level based on past behavior within a fixed observation window, making the system suitable for deployment in near-real-time detection scenarios.}

{The system functions as a behavioral detection framework by combining graph-based learning with implicit temporal and behavioral features. This enables it to flag previously unseen phishing addresses based on their behavior over time and within the network structure. This approach is particularly effective at detecting emerging or stealthy threats that do not match the known phishing profiles but still exhibit suspicious behavior.}

\subsection{Threat Model}

{This work assumes a threat model where the attacker is a regular participant in the Ethereum network with no privileged access. The adversary cannot see system internals or influence the detection model directly. Instead, the attacker operates by creating and controlling multiple addresses, all of which interact with the blockchain through normal transactions.}

{The attacker's main goal is to carry out phishing campaigns by tricking users into sending funds to addresses under the attacker's control. To avoid detection, the attacker may try to mimic benign transaction behavior. For example, they may send low-value transactions, choose normal gas prices, or maintain long idle periods before starting activity. They can also spread activity across several addresses to make their actions less noticeable.}

{However, the detection system is designed to capture behavioral patterns that are difficult to fake. The system relies on implicit features, such as transaction frequency, average time between transactions, gas usage patterns, and transaction timing variability. That reflects how addresses behave over time. These features are based on the statistical behavior observed across historical activity. As a result, they are harder for attackers to manipulate without leaving detectable traces. For instance, an attacker might register a new address and leave it inactive for several days. Then, during early morning hours (\eg, 3:00–4:00 AM UTC), the address suddenly begins to send multiple small transactions to unfamiliar addresses. The attacker increases the gas price to prioritize these transactions. While each individual action might appear normal, the system detects the combination of unusual timing, abrupt change in behavior, and tightly spaced transactions as a suspicious pattern.}

{Because the model learns from a wide set of implicit features and considers the address’s historical and relational behavior in the graph, it can flag such addresses as phishing, even if the attacker tries to blend in. The use of graph-based learning further strengthens detection by aggregating information from neighboring nodes, making it harder for the attacker to isolate their actions.}

{In summary, the assumed adversary is capable of imitating surface-level activity, but cannot easily avoid exposing behavioral inconsistencies. The system’s use of implicit temporal and structural features enables it to detect phishing strategies that would be missed by methods relying only on explicit or static information.}

\section{Experimental setup}\label{sec:experment}

This section outlines the steps taken to evaluate the performance of the GCN model in detecting phishing activities on the Ethereum blockchain. The experiment was divided into three main phases: (1) data preparation, (2) model training and testing, and (3) performance evaluation. The goal of the experiment was to test the efficacy of explicit versus implicit feature sets in distinguishing between phishing and benign transactions. This comparative analysis aims to identify the most impactful features and refine the model’s predictive capabilities for real-world applications.

\subsection{Data Preparation}

\BfPara{Data Collection} Ethereum transaction data was collected using the Etherscan API, comprising 671,865 phishing transactions from 7,915 phishing addresses and 2,687,460 benign transactions. This resulted in a final dataset of 3,359,325 transactions. The comprehensive dataset allows for a robust analysis of phishing patterns and aids in the development of a nuanced understanding of transaction behaviors.

\BfPara{Data Cleaning and Labeling} The dataset was thoroughly cleaned by removing duplicates, null values, and irrelevant transactions. Each transaction was carefully labeled as either phishing or benign. The label distribution was as follows: around 600K phishing and around 2M benign transactions. This step ensures the accuracy and reliability of the training and testing datasets, providing a solid foundation for the subsequent machine learning tasks. The final dataset included important features such as Block Number, Timestamp, Transaction Hash, Sender Address, Receiver Address, Transaction Value, Gas Used, Gas Price, and the phishing or benign label.

\BfPara{Data Splitting} The dataset, consisting of phishing and benign addresses, was divided into training and testing using an 80/20 split to ensure proportional representation and effective model training. This split strategy supports the model’s ability to generalize well to new, unseen data while ensuring that it is robustly trained on a substantial portion of the available data.  The specifics of the split are shown in the~\autoref{tab:dataset_splits}. 


\begin{table}[t] 
\centering 
\caption{Training and testing dataset sizes.} 
\label{tab:dataset_splits}\vspace{-2mm}
\scalebox{0.9}{
\begin{tabular}{lrrr} 
\Xhline{2\arrayrulewidth}
\textbf{Dataset}     & \textbf{Phishing Nodes} & \textbf{Benign Nodes} & \textbf{Edges} \\
\Xhline{1\arrayrulewidth}
Training Set         & 537{,}492               & 2{,}149{,}968          & 1{,}234{,}355  \\
Testing Set          & 134{,}373               & 537{,}492              & 72{,}848       \\
\Xhline{1\arrayrulewidth}
\textbf{Total}       & \textbf{671{,}865}      & \textbf{2{,}687{,}460} & \textbf{1{,}307{,}203} \\
\Xhline{2\arrayrulewidth}
\end{tabular}}
\end{table}

\BfPara{Statistics} The training set contains around 500K phishing transactions and around 2M  benign transactions, connected by around 1M transaction edges. This large number of edges facilitates a detailed network analysis, enhancing the model’s ability to learn from complex relational data. The testing set comprises around 100K phishing transactions and around 500K benign transactions, with 70K edges.

\BfPara{Class Balancing} We employed class weighting during training due to the significant class imbalance, where phishing nodes make up a small portion of the overall dataset. This method adjusts the model’s focus, ensuring that less frequent but critical phishing cases are not overlooked. This approach assigns higher weights to phishing nodes, ensuring the model pays more attention to identifying phishing addresses.

\BfPara{Normalization} Min-Max scaling was applied to the explicit features (\eg transaction value, gas used, gas price), normalizing their values between 0 and 1. This preprocessing step ensures that all features contribute equally to the training process, preventing any single feature from dominating due to differences in scale.

\subsection{Feature Engineering}

Both explicit and implicit features were tested separately to determine their individual contributions to enhancing phishing detection performance. Initially, the model was evaluated using only explicit features to understand their baseline. Subsequently, implicit features were applied in isolation to assess any improvements or changes in detection capabilities. This separate evaluation allows us to precisely compare the impacts of each type of feature on the model’s performance.

\paragraph{Implicit Features} Various implicit features were extracted from Ethereum transaction data to characterize each node's behavior within the transaction network. The following key features were derived from the raw transaction data:

\begin{enumerate}[leftmargin=*]
\item {\bf Transaction Frequency.} The number of transactions initiated by a node (transactions initiated) and the number of transactions received by a node (transactions received) are closely monitored. This measure helps to identify nodes with unusually high or low activity, which can be clearly indicative of either central hubs in legitimate operations or potential points of compromise in fraudulent schemes.

\item {\bf Total Transaction Value.} The total Ether value sent and received by each node (total value sent, total value received). Monitoring the flow of significant sums can help flag nodes that are potentially involved in money laundering or the unauthorized transfer of funds as part of phishing scams.

\item {\bf Gas Usage.} The average gas used by each node for sending and receiving transactions (average gas used for sending, average gas used for receiving) is recorded. Patterns in gas usage can provide vital clues about nodes prioritizing transactions to strategically facilitate quick settlements, a common tactic in phishing to effectively avoid detection.

\item {\bf Time-Based Features} Time-related patterns were captured by computing the mean and standard deviation of transaction hours for sent and received transactions. These metrics include the mean transaction hour for sending, the standard deviation of transaction hour for sending, the mean transaction hour for receiving, and the standard deviation of transaction hour for receiving. By analyzing these figures, we can discern not only the typical hours during which users are most active but also the variability in their transaction times, indicating their temporal transaction habits. This helps in understanding the regularity or randomness of user activities in terms of time, facilitating insights into user behavior and system usage trends.

\item {\bf Transaction Duration.} The time difference between a node's first and last transaction (duration), as well as the average, minimum, and maximum time intervals between transactions (average time between transactions, minimum time between transactions, maximum time between transactions) were calculated to capture temporal transaction patterns. Examining the frequency and regularity of transactions over time allows for the detection of irregular patterns that deviate from normal user behavior, often associated with phishing.

\item {\bf Weekend Transaction Ratio.} The proportion of transactions initiated or received during weekends (weekend transaction ratio for sending, weekend transaction ratio for receiving) was extracted, as phishing addresses may follow distinct temporal patterns compared to benign addresses. By determining the ratio of weekend transactions, it is possible to identify anomalies or consistent trends that clearly differentiate suspicious activities from normal behaviors. This metric helps pinpoint deviations in transactional behavior typically unseen during the regular working week, thus providing a clearer perspective on potentially malicious operations.

\end{enumerate}

To capture more complex interactions and patterns within the transaction network that may not be directly evident from explicit features, graph-based techniques were employed to extract implicit features. Specifically, GCN was applied to the Ethereum transaction graph, where nodes represent Ethereum addresses and edges represent transactions. Through the GCN layers, each node's embedding was iteratively updated by aggregating information from its neighboring nodes, capturing the structural and relational properties of the graph. These implicit features reflect the deeper, latent patterns of node connectivity and behavior within the network, which are critical in distinguishing phishing nodes from benign ones. This method enhances the model’s capability to discern complex patterns of interaction within the Ethereum transaction graph, improving its effectiveness in detecting and addressing potential phishing threats based on network behavior.

\section{Results and Analysis}\label{sec:results}

In this section, we present the GCN's performance when trained and evaluated on explicit and implicit features. We assess the model’s effectiveness using key metrics such as precision, recall, and F1-score, focusing on distinguishing phishing from benign nodes in the Ethereum network. To identify the most distinguishing features, we employ statistical analysis and RF classifier.

\subsection{Distinguishing Features}

We conducted a detailed statistical analysis of key features in transaction data to accurately differentiate phishing from benign nodes. \autoref{tab:key_features} summarizes the features most indicative of phishing behavior, providing insights into the patterns distinguishing these nodes.

\BfPara{Characteristics} We found that the phishing nodes send significantly higher transaction amounts than benign nodes, with an average of $8.27 \times 10^{19}$ compared to $3.72 \times 10^{19}$. Despite similar Max values, phishing nodes show greater variability, making this a key differentiating feature. Phishing nodes also receive more value, averaging $3.88 \times 10^{19}$, compared to $1.73 \times 10^{19}$ for benign nodes. While benign has higher Max values, phishing exhibits more consistent high-value behavior.

Although benign nodes generally use more gas on average (73,375 vs. 44,467 for phishing nodes), phishing nodes display higher outliers, with Max values reaching $2.52 \times 10^{7}$, clearly indicating sporadic spikes in gas usage. Phishing nodes exhibit longer intervals between transactions, averaging $1.71 \times 10^{5}$ seconds, while benign nodes average $-5.16 \times 10^{4}$ seconds. Phishing nodes also show significantly higher variance in the hours they receive transactions (1.00 vs. 0.37 for benign nodes), reflecting more irregular and unpredictable behavior.

The most significant differences between phishing and benign nodes lie notably in transaction volume and timing. Phishing nodes send and receive considerably larger amounts, show higher variability in gas usage, and typically have longer intervals between transactions, making these key indicators essential for distinguishing them from benign nodes.

\begin{table}[t]
\centering
\caption{Comparison of implicit features for phishing vs. benign nodes in terms of mean, max, and standard deviation (Std). Highlighted cells indicate notably higher phishing statistics.}
\label{tab:key_features}
\scalebox{0.85}{
\begin{tabular}{lrrrrrrr}
\Xhline{2\arrayrulewidth}
\multirow{2}{*}{\textbf{Feature}} & \multicolumn{3}{c}{\textbf{Phishing Nodes}} & \multicolumn{3}{c}{\textbf{Benign Nodes}} \\
\cline{2-4} \cline{5-7}
& \textbf{Mean} & \textbf{Max} & \textbf{Std} & \textbf{Mean} & \textbf{Max} & \textbf{Std} \\
\Xhline{1\arrayrulewidth}
G2/G3                        & 5.20            & 20{,}393           & 107.65          & 3.40            & 20{,}393           & 64.53           \\
G4 (Sent)                   & \cellcolor{blue!20}$8.27 \times 10^{19}$ & $4.61 \times 10^{24}$ & \cellcolor{blue!20}$1.26 \times 10^{22}$ & $3.73 \times 10^{19}$ & $4.61 \times 10^{24}$ & $7.09 \times 10^{21}$ \\
G4 (Received)               & \cellcolor{blue!20}$3.88 \times 10^{19}$ & $1.55 \times 10^{24}$ & \cellcolor{blue!20}$4.52 \times 10^{21}$ & $1.74 \times 10^{19}$ & $2.33 \times 10^{24}$ & $3.65 \times 10^{21}$ \\
G7 (Sent)                   & 44{,}467.31      & \cellcolor{blue!20}25{,}258{,}280 & 149{,}682.80     & \cellcolor{blue!20}73{,}375.15 & 7{,}600{,}027      & 120{,}198.40     \\
G7 (Received)               & 1{,}289.49       & 89{,}612           & 5{,}111.78       & 782.73          & 89{,}612           & 4{,}245.39       \\
G1 (Mean hour sent)         & 11.99           & 23                & 6.08            & 11.63           & 23                & 6.01            \\
G1 (Std hour sent)          & 7.99            & 264.50            & 23.77           & 10.71           & 264.50            & 27.94           \\
G1 (Std hour received)      & \cellcolor{blue!20}1.00             & 264.50            & \cellcolor{blue!20}8.01            & 0.38            & 264.50            & 5.24            \\
G1 (Avg time between tx)    & \cellcolor{blue!20}171{,}726.90     & 222{,}295{,}100     & 7{,}434{,}316      & -51{,}675.76     & \cellcolor{blue!20}270{,}628{,}700     & 10{,}037{,}440     \\
G1 (Weekend tx ratio sent)  & 0.29            & 1                 & 0.42            & 0.24            & 1                 & 0.39            \\
G1 (Weekend tx ratio recvd) & 0.02            & 1                 & 0.11            & 0.01            & 1                 & 0.09            \\
\Xhline{2\arrayrulewidth}
\end{tabular}
}
\end{table}

\begin{figure}[t]
    \centering    
    \begin{tikzpicture}
        \begin{axis}[
            width=7.5cm, height=5cm,
            title={},
            xlabel={},
            ylabel={},
            yticklabels={
                {Std Hour Sent},
                {Weekend Transaction Ratio Sent},
                {From Transaction Count},
                {Transaction Duration},
                {Max Time Between Transactions},
                {Avg Time Between Transactions},
                {Mean Hour Sent},
                {Min Time Between Transactions},
                {Avg Gas Used Sent},
                {Total Value Sent}
            },
            ytick=data,
            xbar,
            bar width=5pt,
            enlarge y limits=0.05,
            xmin=0,
            xmax=0.5, 
            xmajorgrids,
            ymajorgrids,
            grid style=dashed,
            tickwidth=0pt,
            axis lines*=left,
            xlabel style={font=\small},
            ylabel style={font=\small},
            xticklabel style={font=\small},
            yticklabel style={font=\small}
        ]
        \addplot coordinates {
            (0.01,0)
            (0.01,1)
            (0.02,2)
            (0.03,3)
            (0.05,4)
            (0.07,5)
            (0.1,6)
            (0.15,7)
            (0.25,8)
            (0.4,9)
        };
        \end{axis}
    \end{tikzpicture}
    \caption{Top 10 important features based on Random Forest.}

    \label{fig:IF-RF}
\end{figure}
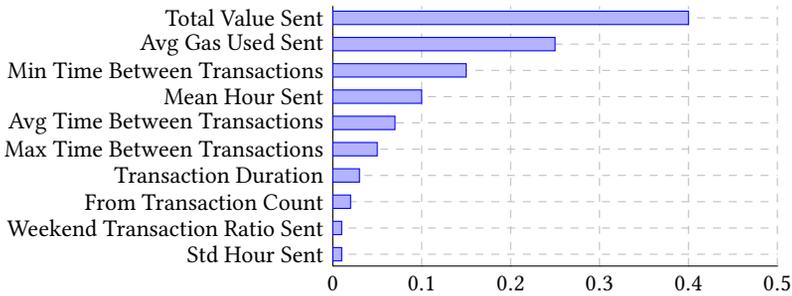

\BfPara{Random Forest Classifier for Feature Importance} RF classifier was applied to accurately quantify the most important features for distinguishing phishing from benign nodes. The RF analysis identified the total value sent as the most important feature (score of 0.40), aligning perfectly with the statistical analysis. The average gas used when sending transactions (score of 0.31) also emerged as a critical feature. Timing-related features such as the average time between transactions further supported the classification of phishing nodes. These feature importance scores are represented in~\autoref{fig:IF-RF}, {which highlights the top 10 most influential implicit features for distinguishing phishing behavior, as ranked by the RF model.}



\takeaway{The combination of high transaction volumes and irregular gas usage and timing provides the strongest basis for distinguishing phishing nodes. These findings highlight the value of combining statistical analysis with machine learning to identify suspicious behavior in Ethereum networks.}

\subsection{Performance on Explicit Features} 

In our first experiment, we trained the GCN using only explicit features, which included basic transactional details such as value, gas used, and timestamps. The model achieved an overall accuracy of 0.79, showing reasonable success in classifying benign nodes. However, its performance in detecting phishing nodes was poor, as reflected by a recall of zero, indicating a complete failure to correctly identify phishing. \autoref{tab:explicit-result} presents the performance metrics.

The low recall for phishing nodes suggests that relying on explicit features limits the model’s ability to capture the nuanced and coordinated behaviors that characterize phishing activities. While the model performed well in classifying benign nodes with a precision of 0.79, it struggled with phishing detection, as evidenced by a precision of only 0.76 and an F1-score of 0.01 for phishing nodes. This result highlights the need for more complex features that can better distinguish between benign and phishing behaviors.

\begin{table}[t]
\centering
\caption{Performance of GCN model using explicit for Phishing and Benign Transactions.}
\label{tab:explicit-result}
\scalebox{0.8}{
\begin{tabular}{lrrrr}
\hline
{Metric}   & {Benign} & {Phishing}  & {Weighted Avg} \\ \hline

{Precision} & \Chart{0.79}     & \Chart{0.76}  & \Chart{0.79}    \\ \hline
{Recall}    & \Chart{1.00}     & \Chart{0.00}  & \Chart{0.79}    \\ \hline
{F1-Score}  & \Chart{0.89}     & \Chart{0.01}  & \Chart{0.70}    \\ \hline
\end{tabular}}
\end{table}

\subsection{Performance on Implicit Features}

The second experiment expanded the feature set by incorporating implicit features derived from the transaction graph. These features capture more complex relational and behavioral patterns, such as transaction frequency and node interactions, providing a richer representation of the Ethereum network's transactional dynamics. When implicit features were included, the model’s overall accuracy improved significantly to 0.95, and the recall for phishing nodes increased to 0.33, (see \autoref{tab:implicit-result}). This improvement indicates that implicit, graph-based features are more effective at capturing the underlying behaviors associated with phishing activities, which often involve coordinated and rapid transactions between nodes. 

\begin{table}[t]
\centering
\caption{Performance of GCN model using implicit for Phishing and Benign Transactions.}
\label{tab:implicit-result}
\scalebox{0.8}{
\begin{tabular}{lrrrr}
\hline
{Metric}   & {Benign} & {Phishing} & {Weighted Avg} \\ \hline
{Precision} & \Chart{0.98}           & \Chart{0.25}                          & \Chart{0.96}                 \\ \hline
{Recall}    & \Chart{0.97}           & \Chart{0.33}                        & \Chart{0.95}               \\ \hline
{F1-Score}  & \Chart{0.97}          & \Chart{0.28}                         & \Chart{0.95}                \\ \hline
\end{tabular}}
\end{table}

The classification report for implicit features underscores the effectiveness of incorporating network-level information into the model. Precision for benign nodes remained high at 0.98, while phishing node precision, although still low at 0.25, shows a marked improvement compared to the experiment with explicit features. The F1-score for phishing nodes increased to 0.28, reflecting a better balance between precision and recall, and further demonstrating the utility of implicit features in detecting phishing activities.

\takeaway{The results of both experiments highlight the limitations of explicit features and the advantages of incorporating implicit, graph-based features. While phishing detection remains challenging, the improvements with implicit features offer promising directions for future detection on blockchain.}


\subsection{Comparative Evaluation}

{To evaluate our approach, we compare our model against recent phishing detection systems as summarized in~\autoref{tab:RW}. These baselines traverse various machine learning, feature types, and experimental settings, enabling a multi-faceted assessment of our method's effectiveness.}

\BfPara{Different Learning Mechanisms} Our method is based on GCN, which allows learning over Ethereum’s transaction graph. In contrast, many prior works adopt alternative learning mechanisms, such as classical machine learning (\eg, decision trees, random forests, XGBoost) or deep learning on tabular data. For example, Kabla~\etal~\cite{KablaAMK22} used ML classifiers over explicit transactional features and reported high F1-scores (0.98) on a relatively small dataset of 5,448 phishing instances. Similarly, Cheng~\etal~\cite{ChengZWLL24} explored hybrid models using LSTM and ML components. However, these approaches rely on limited features and datasets, with weaker generalization in large or imbalanced settings. Our GCN model, by comparison, achieved an F1-score of 0.95 and precision of 0.96 on over 600K phishing cases, demonstrating better performance and scalability in learning from structural context.

\BfPara{Different Feature Types} The majority of prior studies fall into two camps: those using explicit features (\eg, transaction value, timestamp, gas) and those using implicit, graph-derived features (\eg, connectivity, behavior patterns). Zhou~\etal~\cite{Zhou2023} and Kabla~\etal~\cite{KablaAMK22} achieved strong performance with explicit features and ML models, but their datasets were significantly smaller and less diverse. Conversely, works like Li~\etal~\cite{LiGL0LXX23} and Wu~\etal~\cite{WuYLYCCZ22} leveraged implicit features in graph-based or ML-based models. Our study builds upon the latter category but further refines implicit features using temporal patterns and behavioral distributions, such as transaction inter-arrival times and weekend activity ratios. This richer representation contributes to our higher recall (0.95) compared to others, such as Chen~\etal~\cite{ChenPLLXZ21} (recall: 0.14) and Li~\etal~\cite{LiGL0LXX23} (recall: 0.92).

\BfPara{Same Feature Class, Different Instantiations} Among studies utilizing implicit features, our work distinguishes itself through the granularity and temporal depth of the extracted features. Prior studies such as Li~\etal~\cite{LiGL0LXX23} relied on node- or subgraph-level embeddings using contrastive learning, while Fu~\etal~\cite{FuYF22} explored address linkages without temporal decomposition. In contrast, our implicit features were handcrafted to reflect detailed transaction behavior over time. These include statistical metrics such as average, minimum, and maximum inter-transaction intervals; gas usage patterns across sending and receiving behaviors; and distributional metrics like the proportion of weekend activity. We also analyzed behavioral rhythms through time-of-day statistics (mean and standard deviation of transaction hours), providing temporal signatures that distinguish phishing nodes from benign ones. Importantly, we combined this with statistical validation and feature importance ranking using Random Forests to systematically identify and prioritize the most predictive attributes.
These refined implicit features enabled our GCN to capture nuanced behavioral patterns that are often subtle or obscured in generic graph representations. As shown in~\autoref{fig:IF-RF}, the most influential features include total value sent, average gas used when sending, and time-based metrics like minimum time between transactions and standard deviation of send hours. These distinctions set our model apart from other graph-based systems using implicit features but less expressive or temporally coarse representations.

\BfPara{Dataset Scale and Generalization} Our dataset includes over 671K phishing and 2.6M benign transactions, making it one of the largest used in this domain. Many prior works use fewer than 10K phishing samples. Despite scale and class imbalance, our model maintains high precision, recall, and F1-score, showing strong generalization.

\section{Discussion}\label{sec:discussion}

\BfPara{Performance of Explicit Features} In the initial experiment, where only explicit transactional features such as transaction value, gas usage, and timestamps were used, the model exhibited limited ability to distinguish phishing from benign nodes. While these features effectively identified benign transactions, they fell short of capturing the intricate behaviors typical of phishing scams, resulting in a high rate of false negatives. This observation suggests that while explicit features can identify clear-cut cases of normal transactions, they do not provide the nuanced understanding required to detect more deceptive phishing strategies. As such, this highlights the limitations of relying solely on explicit data for detecting phishing, as these features offer only a surface-level view of transaction patterns.

\BfPara{Impact of Implicit Features} When the GCN was tested with implicit features derived from the Ethereum transaction graph, a marked improvement in phishing detection was observed. Implicit features, which capture the relationships and transactional dynamics between nodes, enabled the model to better identify the complex, coordinated activities common in phishing scams. This improvement in detection capability illustrates the critical role of network context in identifying fraud, which is often missed when analyzing transactions in isolation. These graph-based features allowed the model to consider transaction patterns. This underscores the value of focusing on the broader network structure rather than isolated transaction details. However, despite these improvements, the recall for phishing nodes remained at 0.33, indicating that a significant proportion of phishing activities went undetected due to the inherent complexity of phishing patterns, which our current model struggles to fully capture. Graph-based models like GCNs generally require more computational resources compared to traditional machine learning methods, but this cost is often justified by their improved ability to capture complex relational patterns.

\BfPara{Feature Comparison and Model Insights} The comparison between the two experiments demonstrates the distinct roles of explicit and implicit features in phishing detection. While explicit features are useful for understanding basic transactional properties, they lack the depth needed to capture sophisticated behaviors. In contrast, implicit features, which account for interactions and transaction sequences within the network, offer a more detection capability. The GCN’s performance with implicit features highlights the importance of relational data for detecting phishing activities.

\BfPara{Addressing Class Imbalance} A key challenge throughout the experiments was the significant class imbalance, with phishing transactions being vastly outnumbered by benign ones. To mitigate this, a weighted loss function was employed, which improved recall for phishing nodes, especially in the experiment using implicit features. However, despite these improvements, the model still faced issues with false positives, suggesting that further refinement is needed to enhance precision while maintaining high recall. 

\BfPara{Clarifying Performance Comparisons} We understand that comparing our results to prior studies can be misleading if the datasets or evaluation methods differ. In Section~\autoref{sec:results}, we only included these comparisons to give general context, not as a direct benchmark. Our model was trained and tested on a much larger dataset that we built independently, and we reported both per-class and weighted metrics to give a full picture of the model’s performance. Still, we agree that detecting phishing is the most important part, and we have been transparent about the limitations of our recall. In the future, we plan to use shared datasets or re-run baseline models on our data to allow for more consistent and fair comparisons.

\BfPara{Limitations and Future Directions} Although the use of implicit features improved phishing detection, the model’s precision for phishing nodes remains lower than desired, indicating room for optimization. Additionally, the limitations of explicit features highlight the need for approaches to feature engineering. To address the challenge of detecting a higher proportion of phishing, future efforts will focus on refining the model to improve recall, ensuring fewer malicious transactions are missed while maintaining precision.  Additionally, future work could explore integrating advanced techniques such as attention mechanisms or temporal graph networks to further refine phishing detection capabilities. Addressing these challenges could enhance both precision and robustness, especially in real world with imbalanced data. Another important direction is adapting the model to work with streaming Ethereum transactions, allowing it to support real-time phishing detection as transactions occur. Exploring these new methodologies could provide the breakthrough needed to advance the state of phishing detection on the Ethereum blockchain.  We also plan to explore adversarial training to better capture subtle phishing behaviors and improve recall.

\section{Conclusion}\label{sec:conclusion}
This study investigated phishing detection on the Ethereum blockchain using GCNs with explicit and implicit features. While explicit features like value and gas usage provided a basic foundation, they were insufficient for detecting the complex behaviors of phishing attacks. Implicit, graph-based features significantly improved detection by capturing relationships between addresses and broader network patterns. Addressing class imbalance with a weighted loss function enhanced the model’s recall for phishing nodes, but challenges with precision remain, indicating a need for further refinement. Our results highlight the importance of using implicit features for more robust phishing detection. Future work should focus on improving precision and exploring advanced techniques like attention mechanisms and temporal graph networks. In summary, implicit features are essential for detecting phishing activities, and addressing class imbalance will be key to developing more effective detection on blockchain networks.



\begin{thebibliography}{00}


\ifx \showCODEN    \undefined \def \showCODEN     #1{\unskip}     \fi
\ifx \showDOI      \undefined \def \showDOI       #1{#1}\fi
\ifx \showISBNx    \undefined \def \showISBNx     #1{\unskip}     \fi
\ifx \showISBNxiii \undefined \def \showISBNxiii  #1{\unskip}     \fi
\ifx \showISSN     \undefined \def \showISSN      #1{\unskip}     \fi
\ifx \showLCCN     \undefined \def \showLCCN      #1{\unskip}     \fi
\ifx \shownote     \undefined \def \shownote      #1{#1}          \fi
\ifx \showarticletitle \undefined \def \showarticletitle #1{#1}   \fi
\ifx \showURL      \undefined \def \showURL       {\relax}        \fi
\providecommand\bibfield[2]{#2}
\providecommand\bibinfo[2]{#2}
\providecommand\natexlab[1]{#1}
\providecommand\showeprint[2][]{arXiv:#2}

\bibitem[\protect\citeauthoryear{Adeniran, Human, and Mohaisen}{Adeniran
  et~al\mbox{.}}{2024}]%
        {AdeniranHM24}
\bibfield{author}{\bibinfo{person}{Ayodeji Adeniran}, \bibinfo{person}{Kieran
  Human}, {and} \bibinfo{person}{David Mohaisen}.}
  \bibinfo{year}{2024}\natexlab{}.
\newblock \showarticletitle{Dissecting the Infrastructure Used in Web-based
  Cryptojacking: {A} Measurement Perspective}.
\newblock \bibinfo{journal}{{\em CoRR\/}}  \bibinfo{volume}{abs/2408.03426}
  (\bibinfo{year}{2024}).
\newblock
\showDOI{%
\url{https://doi.org/10.48550/ARXIV.2408.03426}}


\bibitem[\protect\citeauthoryear{Ahmad, Saad, Bassiouni, and Mohaisen}{Ahmad
  et~al\mbox{.}}{2018}]%
        {AhmadSBM18}
\bibfield{author}{\bibinfo{person}{Ashar Ahmad}, \bibinfo{person}{Muhammad
  Saad}, \bibinfo{person}{Mostafa~A. Bassiouni}, {and} \bibinfo{person}{Aziz
  Mohaisen}.} \bibinfo{year}{2018}\natexlab{}.
\newblock \showarticletitle{Towards Blockchain-Driven, Secure and Transparent
  Audit Logs}. In \bibinfo{booktitle}{{\em Proceedings of the 15th {EAI}
  International Conference on Mobile and Ubiquitous Systems: Computing,
  Networking and Services, MobiQuitous 2018, 5-7 November 2018, New York City,
  NY, {USA}}}. \bibinfo{publisher}{{ACM}}, \bibinfo{pages}{443--448}.
\newblock
\showDOI{%
\url{https://doi.org/10.1145/3286978.3286985}}


\bibitem[\protect\citeauthoryear{Alghuried and Mohaisen}{Alghuried and
  Mohaisen}{2024}]%
        {AlghuriedM24}
\bibfield{author}{\bibinfo{person}{Ahod Alghuried} {and} \bibinfo{person}{David
  Mohaisen}.} \bibinfo{year}{2024}\natexlab{}.
\newblock
  \showarticletitle{\href{https://doi.org/10.48550/arXiv.2408.03441}{Simple
  Perturbations Subvert Ethereum Phishing Transactions Detection: An Empirical
  Analysis}}. In \bibinfo{booktitle}{{\em The 25th International Conference
  Information Security Applications, {WISA}}}. \bibinfo{pages}{1--12}.
\newblock


\bibitem[\protect\citeauthoryear{Bartoletti, Lande, Loddo, Pompianu, and
  Serusi}{Bartoletti et~al\mbox{.}}{2021}]%
        {BartolettiLLPS21}
\bibfield{author}{\bibinfo{person}{Massimo Bartoletti},
  \bibinfo{person}{Stefano Lande}, \bibinfo{person}{Andrea Loddo},
  \bibinfo{person}{Livio Pompianu}, {and} \bibinfo{person}{Sergio Serusi}.}
  \bibinfo{year}{2021}\natexlab{}.
\newblock \showarticletitle{Cryptocurrency Scams: Analysis and Perspectives}.
\newblock \bibinfo{journal}{{\em {IEEE} Access\/}}  \bibinfo{volume}{9}
  (\bibinfo{year}{2021}), \bibinfo{pages}{148353--148373}.
\newblock
\showDOI{%
\url{https://doi.org/10.1109/ACCESS.2021.3123894}}


\bibitem[\protect\citeauthoryear{Bibi}{Bibi}{2019}]%
        {Bibi19}
\bibfield{author}{\bibinfo{person}{Shaista Bibi}.}
  \bibinfo{year}{2019}\natexlab{}.
\newblock
  \showarticletitle{\href{https://doi.org/10.1145/3297280.3297644}{Cryptocurrency
  world identification and public concerns detection via social media: student
  research abstract}}. In \bibinfo{booktitle}{{\em Proceedings of the 34th
  {ACM/SIGAPP} Symposium on Applied Computing, {SAC}}}.
  \bibinfo{publisher}{{ACM}}, \bibinfo{pages}{550--552}.
\newblock


\bibitem[\protect\citeauthoryear{Cernera, Morgia, Mei, and Sassi}{Cernera
  et~al\mbox{.}}{2023}]%
        {CerneraMMS23}
\bibfield{author}{\bibinfo{person}{Federico Cernera},
  \bibinfo{person}{Massimo~La Morgia}, \bibinfo{person}{Alessandro Mei}, {and}
  \bibinfo{person}{Francesco Sassi}.} \bibinfo{year}{2023}\natexlab{}.
\newblock
  \showarticletitle{\href{https://www.usenix.org/conference/usenixsecurity23/presentation/cernera}{Token
  Spammers, Rug Pulls, and Sniper Bots: An Analysis of the Ecosystem of Tokens
  in Ethereum and in the Binance Smart Chain (BNB)}}. In
  \bibinfo{booktitle}{{\em 32nd {USENIX} Security Symposium}}.
  \bibinfo{publisher}{{USENIX} Association}, \bibinfo{pages}{3349--3366}.
\newblock


\bibitem[\protect\citeauthoryear{Chen, Pendleton, Njilla, and Xu}{Chen
  et~al\mbox{.}}{2021a}]%
        {ChenPNX20}
\bibfield{author}{\bibinfo{person}{Huashan Chen}, \bibinfo{person}{Marcus
  Pendleton}, \bibinfo{person}{Laurent Njilla}, {and} \bibinfo{person}{Shouhuai
  Xu}.} \bibinfo{year}{2021}\natexlab{a}.
\newblock \showarticletitle{\href{https://doi.org/10.1145/3391195}{A Survey on
  Ethereum Systems Security: Vulnerabilities, Attacks, and Defenses}}.
\newblock \bibinfo{journal}{{\em {ACM} Comput. Surv.\/}} \bibinfo{volume}{53},
  \bibinfo{number}{3} (\bibinfo{year}{2021}), \bibinfo{pages}{67:1--67:43}.
\newblock


\bibitem[\protect\citeauthoryear{Chen, Peng, Liu, Li, Xie, and Zheng}{Chen
  et~al\mbox{.}}{2021b}]%
        {ChenPLLXZ21}
\bibfield{author}{\bibinfo{person}{Liang Chen}, \bibinfo{person}{Jiaying Peng},
  \bibinfo{person}{Yang Liu}, \bibinfo{person}{Jintang Li},
  \bibinfo{person}{Fenfang Xie}, {and} \bibinfo{person}{Zibin Zheng}.}
  \bibinfo{year}{2021}\natexlab{b}.
\newblock \showarticletitle{\href{https://doi.org/10.1145/3398071}{Phishing
  Scams Detection in Ethereum Transaction Network}}.
\newblock \bibinfo{journal}{{\em {ACM} Trans. Internet Techn.\/}}
  \bibinfo{volume}{21}, \bibinfo{number}{1} (\bibinfo{year}{2021}),
  \bibinfo{pages}{10:1--10:16}.
\newblock


\bibitem[\protect\citeauthoryear{Chen, Guo, Chen, Zheng, and Lu}{Chen
  et~al\mbox{.}}{2020}]%
        {ChenGCZL20}
\bibfield{author}{\bibinfo{person}{Weili Chen}, \bibinfo{person}{Xiongfeng
  Guo}, \bibinfo{person}{Zhiguang Chen}, \bibinfo{person}{Zibin Zheng}, {and}
  \bibinfo{person}{Yutong Lu}.} \bibinfo{year}{2020}\natexlab{}.
\newblock
  \showarticletitle{\href{https://doi.org/10.24963/ijcai.2020/621}{Phishing
  Scam Detection on Ethereum: Towards Financial Security for Blockchain
  Ecosystem}}. In \bibinfo{booktitle}{{\em Proceedings of the Twenty-Ninth
  International Joint Conference on Artificial Intelligence, {IJCAI}}}.
  \bibinfo{publisher}{ijcai.org}, \bibinfo{pages}{4506--4512}.
\newblock


\bibitem[\protect\citeauthoryear{Cheng, Zhu, Wang, Liang, and Liu}{Cheng
  et~al\mbox{.}}{2023}]%
        {Cheng0WLL23}
\bibfield{author}{\bibinfo{person}{Ling Cheng}, \bibinfo{person}{Feida Zhu},
  \bibinfo{person}{Yong Wang}, \bibinfo{person}{Ruicheng Liang}, {and}
  \bibinfo{person}{Huiwen Liu}.} \bibinfo{year}{2023}\natexlab{}.
\newblock
  \showarticletitle{\href{https://doi.org/10.1145/3580305.3599817}{Evolve Path
  Tracer: Early Detection of Malicious Addresses in Cryptocurrency}}. In
  \bibinfo{booktitle}{{\em Proceedings of the 29th {ACM} {SIGKDD} Conference on
  Knowledge Discovery and Data Mining, {KDD}}}. \bibinfo{publisher}{{ACM}},
  \bibinfo{pages}{3889--3900}.
\newblock


\bibitem[\protect\citeauthoryear{Cheng, Zhu, Wang, Liang, and Liu}{Cheng
  et~al\mbox{.}}{2024}]%
        {ChengZWLL24}
\bibfield{author}{\bibinfo{person}{Ling Cheng}, \bibinfo{person}{Feida Zhu},
  \bibinfo{person}{Yong Wang}, \bibinfo{person}{Ruicheng Liang}, {and}
  \bibinfo{person}{Huiwen Liu}.} \bibinfo{year}{2024}\natexlab{}.
\newblock \showarticletitle{\href{https://doi.org/10.1145/3626102}{From Asset
  Flow to Status, Action, and Intention Discovery: Early Malice Detection in
  Cryptocurrency}}.
\newblock \bibinfo{journal}{{\em {ACM} Trans. Knowl. Discov. Data\/}}
  \bibinfo{volume}{18}, \bibinfo{number}{3} (\bibinfo{year}{2024}),
  \bibinfo{pages}{50:1--50:27}.
\newblock


\bibitem[\protect\citeauthoryear{Fu, Yu, and Feng}{Fu et~al\mbox{.}}{2022}]%
        {FuYF22}
\bibfield{author}{\bibinfo{person}{Bingxue Fu}, \bibinfo{person}{Xing Yu},
  {and} \bibinfo{person}{Tao Feng}.} \bibinfo{year}{2022}\natexlab{}.
\newblock
  \showarticletitle{\href{https://doi.org/10.1007/s10207-022-00606-6}{CT-GCN: a
  phishing identification model for blockchain cryptocurrency transactions}}.
\newblock \bibinfo{journal}{{\em Int. J. Inf. Sec.\/}} \bibinfo{volume}{21},
  \bibinfo{number}{6} (\bibinfo{year}{2022}), \bibinfo{pages}{1223--1232}.
\newblock


\bibitem[\protect\citeauthoryear{Galletta and Pinelli}{Galletta and
  Pinelli}{2024}]%
        {GallettaP24}
\bibfield{author}{\bibinfo{person}{Letterio Galletta} {and}
  \bibinfo{person}{Fabio Pinelli}.} \bibinfo{year}{2024}\natexlab{}.
\newblock
  \showarticletitle{\href{https://doi.org/10.1145/3605098.3636060}{Explainable
  Ponzi Schemes Detection on Ethereum}}. In \bibinfo{booktitle}{{\em
  Proceedings of the 39th {ACM/SIGAPP} Symposium on Applied Computing, {SAC}}},
  \bibfield{editor}{\bibinfo{person}{Jiman Hong} {and} \bibinfo{person}{Juw~Won
  Park}} (Eds.). \bibinfo{publisher}{{ACM}}, \bibinfo{pages}{1014--1023}.
\newblock


\bibitem[\protect\citeauthoryear{He, Chen, Chen, Hu, Hu, Wu, Chang, Wang, and
  Zhou}{He et~al\mbox{.}}{2023}]%
        {HeCCHHWCWZ23}
\bibfield{author}{\bibinfo{person}{Bowen He}, \bibinfo{person}{Yuan Chen},
  \bibinfo{person}{Zhuo Chen}, \bibinfo{person}{Xiaohui Hu},
  \bibinfo{person}{Yufeng Hu}, \bibinfo{person}{Lei Wu}, \bibinfo{person}{Rui
  Chang}, \bibinfo{person}{Haoyu Wang}, {and} \bibinfo{person}{Yajin Zhou}.}
  \bibinfo{year}{2023}\natexlab{}.
\newblock
  \showarticletitle{\href{https://doi.org/10.1145/3576915.3623210}{TxPhishScope:
  Towards Detecting and Understanding Transaction-based Phishing on Ethereum}}.
  In \bibinfo{booktitle}{{\em Proceedings of the 2023 {ACM} {SIGSAC} Conference
  on Computer and Communications Security, {CCS}}}. \bibinfo{publisher}{{ACM}},
  \bibinfo{pages}{120--134}.
\newblock


\bibitem[\protect\citeauthoryear{Heo, Woo, Yoon, Kang, and Shin}{Heo
  et~al\mbox{.}}{2023}]%
        {HeoWYKS23}
\bibfield{author}{\bibinfo{person}{Hwanjo Heo}, \bibinfo{person}{Seungwon Woo},
  \bibinfo{person}{Taeung Yoon}, \bibinfo{person}{Min~Suk Kang}, {and}
  \bibinfo{person}{Seungwon Shin}.} \bibinfo{year}{2023}\natexlab{}.
\newblock
  \showarticletitle{\href{https://www.ndss-symposium.org/ndss-paper/partitioning-ethereum-without-eclipsing-it/}{Partitioning
  Ethereum without Eclipsing It}}. In \bibinfo{booktitle}{{\em 30th Annual
  Network and Distributed System Security Symposium, {NDSS}}}.
  \bibinfo{publisher}{The Internet Society}.
\newblock


\bibitem[\protect\citeauthoryear{Hornuf, Momtaz, Nam, and Yuan}{Hornuf
  et~al\mbox{.}}{2023}]%
        {HornufMNY2023}
\bibfield{author}{\bibinfo{person}{Lars Hornuf}, \bibinfo{person}{Paul~P
  Momtaz}, \bibinfo{person}{Rachel~J Nam}, {and} \bibinfo{person}{Ye Yuan}.}
  \bibinfo{year}{2023}\natexlab{}.
\newblock
  \showarticletitle{\href{http://dx.doi.org/10.2139/ssrn.4527415}{Cybercrime on
  the ethereum blockchain}}.
\newblock  (\bibinfo{year}{2023}).
\newblock


\bibitem[\protect\citeauthoryear{Jones, Armstrong, Tornblad, and Namin}{Jones
  et~al\mbox{.}}{2021}]%
        {JonesATN21}
\bibfield{author}{\bibinfo{person}{Keith~S. Jones}, \bibinfo{person}{Miriam~E.
  Armstrong}, \bibinfo{person}{McKenna~K. Tornblad}, {and}
  \bibinfo{person}{Akbar~Siami Namin}.} \bibinfo{year}{2021}\natexlab{}.
\newblock \showarticletitle{How social engineers use persuasion principles
  during vishing attacks}.
\newblock \bibinfo{journal}{{\em Inf. Comput. Secur.\/}} \bibinfo{volume}{29},
  \bibinfo{number}{2} (\bibinfo{year}{2021}), \bibinfo{pages}{314--331}.
\newblock
\showDOI{%
\url{https://doi.org/10.1108/ICS-07-2020-0113}}


\bibitem[\protect\citeauthoryear{Kabla, Anbar, Manickam, and Karuppayah}{Kabla
  et~al\mbox{.}}{2022}]%
        {KablaAMK22}
\bibfield{author}{\bibinfo{person}{Arkan Hammoodi~Hasan Kabla},
  \bibinfo{person}{Mohammed Anbar}, \bibinfo{person}{Selvakumar Manickam},
  {and} \bibinfo{person}{Shankar Karuppayah}.} \bibinfo{year}{2022}\natexlab{}.
\newblock
  \showarticletitle{\href{https://doi.org/10.1109/ACCESS.2022.3220780}{Eth-PSD:
  A Machine Learning-Based Phishing Scam Detection Approach in Ethereum}}.
\newblock \bibinfo{journal}{{\em {IEEE} Access\/}}  \bibinfo{volume}{10}
  (\bibinfo{year}{2022}), \bibinfo{pages}{118043--118057}.
\newblock


\bibitem[\protect\citeauthoryear{Kampers, Qahtan, Mathur, and
  Velegrakis}{Kampers et~al\mbox{.}}{2022}]%
        {KampersQMV22}
\bibfield{author}{\bibinfo{person}{Olaf Kampers},
  \bibinfo{person}{Abdulhakim~Ali Qahtan}, \bibinfo{person}{Swati Mathur},
  {and} \bibinfo{person}{Yannis Velegrakis}.} \bibinfo{year}{2022}\natexlab{}.
\newblock
  \showarticletitle{\href{https://doi.org/10.1145/3477314.3507185}{Manipulation
  detection in cryptocurrency markets: an anomaly and change detection based
  approach}}. In \bibinfo{booktitle}{{\em {SAC} '22: The 37th {ACM/SIGAPP}
  Symposium on Applied Computing}}. \bibinfo{publisher}{{ACM}},
  \bibinfo{pages}{326--329}.
\newblock


\bibitem[\protect\citeauthoryear{Kimber}{Kimber}{2023}]%
        {Kimber2023}
\bibfield{author}{\bibinfo{person}{Author~Name Kimber}.}
  \bibinfo{year}{2023}\natexlab{}.
\newblock \bibinfo{title}{Scams on Ethereum}.
\newblock   (\bibinfo{year}{2023}).
\newblock
\showURL{%
\url{https://github.com/YNclusk/scamsonethereum}}


\bibitem[\protect\citeauthoryear{Li, Gou, Liu, Hou, Li, and Xiong}{Li
  et~al\mbox{.}}{2022}]%
        {LiGLHLX22}
\bibfield{author}{\bibinfo{person}{Sijia Li}, \bibinfo{person}{Gaopeng Gou},
  \bibinfo{person}{Chang Liu}, \bibinfo{person}{Chengshang Hou},
  \bibinfo{person}{Zhenzhen Li}, {and} \bibinfo{person}{Gang Xiong}.}
  \bibinfo{year}{2022}\natexlab{}.
\newblock
  \showarticletitle{\href{https://doi.org/10.1145/3485447.3512226}{TTAGN:
  Temporal Transaction Aggregation Graph Network for Ethereum Phishing Scams
  Detection}}. In \bibinfo{booktitle}{{\em {WWW} '22: The {ACM} Web Conference
  2022}}. \bibinfo{publisher}{{ACM}}, \bibinfo{pages}{661--669}.
\newblock


\bibitem[\protect\citeauthoryear{Li, Gou, Liu, Xiong, Li, Xiao, and Xing}{Li
  et~al\mbox{.}}{2023a}]%
        {LiGL0LXX23}
\bibfield{author}{\bibinfo{person}{Sijia Li}, \bibinfo{person}{Gaopeng Gou},
  \bibinfo{person}{Chang Liu}, \bibinfo{person}{Gang Xiong},
  \bibinfo{person}{Zhen Li}, \bibinfo{person}{Junchao Xiao}, {and}
  \bibinfo{person}{Xinyu Xing}.} \bibinfo{year}{2023}\natexlab{a}.
\newblock \showarticletitle{\href{https://doi.org/10.1145/3627106.3627109}{TGC:
  Transaction Graph Contrast Network for Ethereum Phishing Scam Detection}}. In
  \bibinfo{booktitle}{{\em Annual Computer Security Applications Conference,
  {ACSAC}}}. \bibinfo{publisher}{{ACM}}, \bibinfo{pages}{352--365}.
\newblock


\bibitem[\protect\citeauthoryear{Li, Wang, Wu, Zhong, and Xu}{Li
  et~al\mbox{.}}{2023b}]%
        {LiWW0X23}
\bibfield{author}{\bibinfo{person}{Shucheng Li}, \bibinfo{person}{Runchuan
  Wang}, \bibinfo{person}{Hao Wu}, \bibinfo{person}{Sheng Zhong}, {and}
  \bibinfo{person}{Fengyuan Xu}.} \bibinfo{year}{2023}\natexlab{b}.
\newblock
  \showarticletitle{\href{https://doi.org/10.1145/3581783.3612461}{SIEGE:
  Self-Supervised Incremental Deep Graph Learning for Ethereum Phishing Scam
  Detection}}. In \bibinfo{booktitle}{{\em Proceedings of the 31st {ACM}
  International Conference on Multimedia, {MM}}}. \bibinfo{publisher}{{ACM}},
  \bibinfo{pages}{8881--8890}.
\newblock


\bibitem[\protect\citeauthoryear{Lin, Xiao, Hu, Zhang, Liu, and Luo}{Lin
  et~al\mbox{.}}{2023}]%
        {LinXHZLL23}
\bibfield{author}{\bibinfo{person}{Zhutian Lin}, \bibinfo{person}{Xi Xiao},
  \bibinfo{person}{Guangwu Hu}, \bibinfo{person}{Bin Zhang},
  \bibinfo{person}{Qixu Liu}, {and} \bibinfo{person}{Xiapu Luo}.}
  \bibinfo{year}{2023}\natexlab{}.
\newblock
  \showarticletitle{\href{https://doi.org/10.1109/SECON58729.2023.10287480}{Phish2vec:
  A Temporal and Heterogeneous Network Embedding Approach for Detecting
  Phishing Scams on Ethereum}}. In \bibinfo{booktitle}{{\em 20th Annual {IEEE}
  International Conference on Sensing, Communication, and Networking,
  {SECON}}}. \bibinfo{publisher}{{IEEE}}.
\newblock


\bibitem[\protect\citeauthoryear{Liu, Chen, Wu, Wu, Fang, and Zheng}{Liu
  et~al\mbox{.}}{2024}]%
        {LiuCWWFZ24}
\bibfield{author}{\bibinfo{person}{Jieli Liu}, \bibinfo{person}{Jinze Chen},
  \bibinfo{person}{Jiajing Wu}, \bibinfo{person}{Zhiying Wu},
  \bibinfo{person}{Junyuan Fang}, {and} \bibinfo{person}{Zibin Zheng}.}
  \bibinfo{year}{2024}\natexlab{}.
\newblock
  \showarticletitle{\href{https://doi.org/10.1109/TIFS.2024.3359000}{Fishing
  for Fraudsters: Uncovering Ethereum Phishing Gangs With Blockchain Data}}.
\newblock \bibinfo{journal}{{\em {IEEE} Trans. Inf. Forensics Secur.\/}}
  \bibinfo{volume}{19} (\bibinfo{year}{2024}), \bibinfo{pages}{3038--3050}.
\newblock


\bibitem[\protect\citeauthoryear{{NetworkX}}{{NetworkX}}{2024}]%
        {networkx}
\bibfield{author}{\bibinfo{person}{{NetworkX}}.}
  \bibinfo{year}{2024}\natexlab{}.
\newblock \bibinfo{title}{NetworkX}.
\newblock \bibinfo{howpublished}{\url{https://networkx.org/}}.
  (\bibinfo{year}{2024}).
\newblock
\newblock
\shownote{Accessed: 2024-11.}


\bibitem[\protect\citeauthoryear{Ngo, Winarto, Kou, Park, Akram, and Lee}{Ngo
  et~al\mbox{.}}{2019}]%
        {NgoWKPAL19}
\bibfield{author}{\bibinfo{person}{Cuong~Phuc Ngo},
  \bibinfo{person}{Amadeus~Aristo Winarto}, \bibinfo{person}{Connie Khor~Li
  Kou}, \bibinfo{person}{Sojeong Park}, \bibinfo{person}{Farhan Akram}, {and}
  \bibinfo{person}{Hwee~Kuan Lee}.} \bibinfo{year}{2019}\natexlab{}.
\newblock
  \showarticletitle{\href{https://doi.org/10.1109/ICTAI.2019.00028}{Fence GAN:
  Towards Better Anomaly Detection}}. In \bibinfo{booktitle}{{\em 31st {IEEE}
  International Conference on Tools with Artificial Intelligence, {ICTAI}}}.
  \bibinfo{publisher}{{IEEE}}, \bibinfo{pages}{141--148}.
\newblock


\bibitem[\protect\citeauthoryear{Pinio, Batko, and Lewicka}{Pinio
  et~al\mbox{.}}{2024}]%
        {PinioBL24}
\bibfield{author}{\bibinfo{person}{Pawel Pinio}, \bibinfo{person}{Roman Batko},
  {and} \bibinfo{person}{Dagmara Lewicka}.} \bibinfo{year}{2024}\natexlab{}.
\newblock
  \showarticletitle{\href{https://doi.org/10.1145/3647722.3647729}{Between
  Theory and Value Transactions: A Multifaceted Exploration of Relevance and
  Resilience of Decentralised Autonomous Organisations}}. In
  \bibinfo{booktitle}{{\em Proceedings of the 2024 7th International Conference
  on Software Engineering and Information Management, {ICSIM}}}.
  \bibinfo{publisher}{{ACM}}, \bibinfo{pages}{42--48}.
\newblock


\bibitem[\protect\citeauthoryear{Saad, Ahmad, and Mohaisen}{Saad
  et~al\mbox{.}}{2019}]%
        {SaadAM19}
\bibfield{author}{\bibinfo{person}{Muhammad Saad}, \bibinfo{person}{Ashar
  Ahmad}, {and} \bibinfo{person}{Aziz Mohaisen}.}
  \bibinfo{year}{2019}\natexlab{}.
\newblock \showarticletitle{Fighting Fake News Propagation with Blockchains}.
  In \bibinfo{booktitle}{{\em 7th {IEEE} Conference on Communications and
  Network Security, {CNS} 2019, Washington, DC, USA, June 10-12, 2019}}.
  \bibinfo{publisher}{{IEEE}}, \bibinfo{pages}{1--4}.
\newblock
\showDOI{%
\url{https://doi.org/10.1109/CNS.2019.8802670}}


\bibitem[\protect\citeauthoryear{Saad, Anwar, Ahmad, Alasmary, Yuksel, and
  Mohaisen}{Saad et~al\mbox{.}}{2022}]%
        {SaadAAAYM22}
\bibfield{author}{\bibinfo{person}{Muhammad Saad}, \bibinfo{person}{Afsah
  Anwar}, \bibinfo{person}{Ashar Ahmad}, \bibinfo{person}{Hisham Alasmary},
  \bibinfo{person}{Murat Yuksel}, {and} \bibinfo{person}{David Mohaisen}.}
  \bibinfo{year}{2022}\natexlab{}.
\newblock \showarticletitle{\emph{RouteChain}: Towards Blockchain-based secure
  and efficient {BGP} routing}.
\newblock \bibinfo{journal}{{\em Comput. Networks\/}}  \bibinfo{volume}{217}
  (\bibinfo{year}{2022}), \bibinfo{pages}{109362}.
\newblock
\showDOI{%
\url{https://doi.org/10.1016/J.COMNET.2022.109362}}


\bibitem[\protect\citeauthoryear{Saad, Chen, and Mohaisen}{Saad
  et~al\mbox{.}}{2021}]%
        {SaadCM21}
\bibfield{author}{\bibinfo{person}{Muhammad Saad}, \bibinfo{person}{Songqing
  Chen}, {and} \bibinfo{person}{David Mohaisen}.}
  \bibinfo{year}{2021}\natexlab{}.
\newblock \showarticletitle{SyncAttack: Double-spending in Bitcoin Without
  Mining Power}. In \bibinfo{booktitle}{{\em {CCS} '21: 2021 {ACM} {SIGSAC}
  Conference on Computer and Communications Security, Virtual Event, Republic
  of Korea, November 15 - 19, 2021}}. \bibinfo{publisher}{{ACM}},
  \bibinfo{pages}{1668--1685}.
\newblock
\showDOI{%
\url{https://doi.org/10.1145/3460120.3484568}}


\bibitem[\protect\citeauthoryear{Saad, Choi, Nyang, Kim, and Mohaisen}{Saad
  et~al\mbox{.}}{2020}]%
        {SaadCNKM20}
\bibfield{author}{\bibinfo{person}{Muhammad Saad}, \bibinfo{person}{Jinchun
  Choi}, \bibinfo{person}{DaeHun Nyang}, \bibinfo{person}{Joongheon Kim}, {and}
  \bibinfo{person}{Aziz Mohaisen}.} \bibinfo{year}{2020}\natexlab{}.
\newblock \showarticletitle{Toward Characterizing Blockchain-Based
  Cryptocurrencies for Highly Accurate Predictions}.
\newblock \bibinfo{journal}{{\em {IEEE} Syst. J.\/}} \bibinfo{volume}{14},
  \bibinfo{number}{1} (\bibinfo{year}{2020}), \bibinfo{pages}{321--332}.
\newblock
\showDOI{%
\url{https://doi.org/10.1109/JSYST.2019.2927707}}


\bibitem[\protect\citeauthoryear{Saad, Kim, Nyang, and Mohaisen}{Saad
  et~al\mbox{.}}{2021}]%
        {SaadKNM21}
\bibfield{author}{\bibinfo{person}{Muhammad Saad}, \bibinfo{person}{Joongheon
  Kim}, \bibinfo{person}{DaeHun Nyang}, {and} \bibinfo{person}{David
  Mohaisen}.} \bibinfo{year}{2021}\natexlab{}.
\newblock \showarticletitle{Contra-{\({_\ast}\)}: Mechanisms for countering
  spam attacks on blockchain's memory pools}.
\newblock \bibinfo{journal}{{\em J. Netw. Comput. Appl.\/}}
  \bibinfo{volume}{179} (\bibinfo{year}{2021}), \bibinfo{pages}{102971}.
\newblock
\showDOI{%
\url{https://doi.org/10.1016/J.JNCA.2020.102971}}


\bibitem[\protect\citeauthoryear{Saad and Mohaisen}{Saad and Mohaisen}{2023}]%
        {SaadM23}
\bibfield{author}{\bibinfo{person}{Muhammad Saad} {and} \bibinfo{person}{David
  Mohaisen}.} \bibinfo{year}{2023}\natexlab{}.
\newblock
  \showarticletitle{\href{https://doi.org/10.1109/SP46215.2023.10179456}{Three
  Birds with One Stone: Efficient Partitioning Attacks on Interdependent
  Cryptocurrency Networks}}. In \bibinfo{booktitle}{{\em 44th {IEEE} Symposium
  on Security and Privacy, {SP}}}. \bibinfo{publisher}{{IEEE}},
  \bibinfo{pages}{111--125}.
\newblock


\bibitem[\protect\citeauthoryear{Saad, Njilla, Kamhoua, Kim, Nyang, and
  Mohaisen}{Saad et~al\mbox{.}}{2019b}]%
        {SaadNKKNM19}
\bibfield{author}{\bibinfo{person}{Muhammad Saad}, \bibinfo{person}{Laurent
  Njilla}, \bibinfo{person}{Charles~A. Kamhoua}, \bibinfo{person}{Joongheon
  Kim}, \bibinfo{person}{DaeHun Nyang}, {and} \bibinfo{person}{Aziz Mohaisen}.}
  \bibinfo{year}{2019}\natexlab{b}.
\newblock \showarticletitle{Mempool optimization for Defending Against DDoS
  Attacks in PoW-based Blockchain Systems}. In \bibinfo{booktitle}{{\em {IEEE}
  International Conference on Blockchain and Cryptocurrency, {ICBC} 2019,
  Seoul, Korea (South), May 14-17, 2019}}. \bibinfo{publisher}{{IEEE}},
  \bibinfo{pages}{285--292}.
\newblock
\showDOI{%
\url{https://doi.org/10.1109/BLOC.2019.8751476}}


\bibitem[\protect\citeauthoryear{Saad, Njilla, Kamhoua, and Mohaisen}{Saad
  et~al\mbox{.}}{2019a}]%
        {SaadNKM19}
\bibfield{author}{\bibinfo{person}{Muhammad Saad}, \bibinfo{person}{Laurent
  Njilla}, \bibinfo{person}{Charles~A. Kamhoua}, {and} \bibinfo{person}{Aziz
  Mohaisen}.} \bibinfo{year}{2019}\natexlab{a}.
\newblock \showarticletitle{Countering Selfish Mining in Blockchains}. In
  \bibinfo{booktitle}{{\em International Conference on Computing, Networking
  and Communications, {ICNC} 2019, Honolulu, HI, USA, February 18-21, 2019}}.
  \bibinfo{publisher}{{IEEE}}, \bibinfo{pages}{360--364}.
\newblock
\showDOI{%
\url{https://doi.org/10.1109/ICCNC.2019.8685577}}


\bibitem[\protect\citeauthoryear{Saad, Spaulding, Njilla, Kamhoua, Shetty,
  Nyang, and Mohaisen}{Saad et~al\mbox{.}}{2020}]%
        {SaadSNKSNM20}
\bibfield{author}{\bibinfo{person}{Muhammad Saad}, \bibinfo{person}{Jeffrey
  Spaulding}, \bibinfo{person}{Laurent Njilla}, \bibinfo{person}{Charles~A.
  Kamhoua}, \bibinfo{person}{Sachin Shetty}, \bibinfo{person}{DaeHun Nyang},
  {and} \bibinfo{person}{David Mohaisen}.} \bibinfo{year}{2020}\natexlab{}.
\newblock
  \showarticletitle{\href{https://doi.org/10.1109/COMST.2020.2975999}{Exploring
  the Attack Surface of Blockchain: A Comprehensive Survey}}.
\newblock \bibinfo{journal}{{\em {IEEE} Commun. Surv. Tutorials\/}}
  \bibinfo{volume}{22}, \bibinfo{number}{3} (\bibinfo{year}{2020}),
  \bibinfo{pages}{1977--2008}.
\newblock


\bibitem[\protect\citeauthoryear{Vasek and Moore}{Vasek and Moore}{2015}]%
        {VasekM15}
\bibfield{author}{\bibinfo{person}{Marie Vasek} {and} \bibinfo{person}{Tyler
  Moore}.} \bibinfo{year}{2015}\natexlab{}.
\newblock
  \showarticletitle{\href{https://doi.org/10.1007/978-3-662-47854-7_4}{There's
  No Free Lunch, Even Using Bitcoin: Tracking the Popularity and Profits of
  Virtual Currency Scams}}. In \bibinfo{booktitle}{{\em Financial Cryptography
  and Data Security - 19th International Conference, {FC}}} {\em
  (\bibinfo{series}{Lecture Notes in Computer Science})},
  Vol.~\bibinfo{volume}{8975}. \bibinfo{publisher}{Springer},
  \bibinfo{pages}{44--61}.
\newblock


\bibitem[\protect\citeauthoryear{Wan, Xiao, and Zhang}{Wan
  et~al\mbox{.}}{2023}]%
        {WanXZ23}
\bibfield{author}{\bibinfo{person}{Yun Wan}, \bibinfo{person}{Feng Xiao}, {and}
  \bibinfo{person}{Dapeng Zhang}.} \bibinfo{year}{2023}\natexlab{}.
\newblock \showarticletitle{Early-stage phishing detection on the Ethereum
  transaction network}.
\newblock \bibinfo{journal}{{\em Soft Comput.\/}} \bibinfo{volume}{27},
  \bibinfo{number}{7} (\bibinfo{year}{2023}), \bibinfo{pages}{3707--3719}.
\newblock
\showDOI{%
\url{https://doi.org/10.1007/S00500-022-07661-0}}


\bibitem[\protect\citeauthoryear{Wang, Chen, Xu, Wu, Shen, Xuan, and Yang}{Wang
  et~al\mbox{.}}{2022}]%
        {WangCXWSXY22}
\bibfield{author}{\bibinfo{person}{Jinhuan Wang}, \bibinfo{person}{Pengtao
  Chen}, \bibinfo{person}{Xinyao Xu}, \bibinfo{person}{Jiajing Wu},
  \bibinfo{person}{Meng Shen}, \bibinfo{person}{Qi Xuan}, {and}
  \bibinfo{person}{Xiaoniu Yang}.} \bibinfo{year}{2022}\natexlab{}.
\newblock
  \showarticletitle{\href{https://doi.org/10.48550/arXiv.2208.12938}{TSGN:
  Transaction Subgraph Networks Assisting Phishing Detection in Ethereum}}.
\newblock \bibinfo{journal}{{\em CoRR\/}}  \bibinfo{volume}{abs/2208.12938}
  (\bibinfo{year}{2022}).
\newblock


\bibitem[\protect\citeauthoryear{Wang, Tong, Pang, Wang, and Han}{Wang
  et~al\mbox{.}}{2024}]%
        {WangTPWH24}
\bibfield{author}{\bibinfo{person}{Kai Wang}, \bibinfo{person}{Michael Tong},
  \bibinfo{person}{Jun Pang}, \bibinfo{person}{Jitao Wang}, {and}
  \bibinfo{person}{Weili Han}.} \bibinfo{year}{2024}\natexlab{}.
\newblock \showarticletitle{\href{https://doi.org/10.1145/3687487}{XRAD:
  Ransomware Address Detection Method based on Bitcoin Transaction
  Relationships}}.
\newblock \bibinfo{journal}{{\em {ACM} Trans. Web\/}} (\bibinfo{year}{2024}).
\newblock


\bibitem[\protect\citeauthoryear{Wen, Xiao, Wang, and Wang}{Wen
  et~al\mbox{.}}{2023}]%
        {WenXWW23}
\bibfield{author}{\bibinfo{person}{Tingke Wen}, \bibinfo{person}{Yuanxing
  Xiao}, \bibinfo{person}{Anqi Wang}, {and} \bibinfo{person}{Haizhou Wang}.}
  \bibinfo{year}{2023}\natexlab{}.
\newblock \showarticletitle{\href{https://doi.org/10.1016/j.eswa.2022.118463}{A
  novel hybrid feature fusion model for detecting phishing scam on Ethereum
  using deep neural network}}.
\newblock \bibinfo{journal}{{\em Expert Syst. Appl.\/}}  \bibinfo{volume}{211}
  (\bibinfo{year}{2023}), \bibinfo{pages}{118463}.
\newblock


\bibitem[\protect\citeauthoryear{Wu, Yuan, Lin, You, Chen, Chen, and Zheng}{Wu
  et~al\mbox{.}}{2022}]%
        {WuYLYCCZ22}
\bibfield{author}{\bibinfo{person}{Jiajing Wu}, \bibinfo{person}{Qi Yuan},
  \bibinfo{person}{Dan Lin}, \bibinfo{person}{Wei You}, \bibinfo{person}{Weili
  Chen}, \bibinfo{person}{Chuan Chen}, {and} \bibinfo{person}{Zibin Zheng}.}
  \bibinfo{year}{2022}\natexlab{}.
\newblock
  \showarticletitle{\href{https://doi.org/10.1109/TSMC.2020.3016821}{Who Are
  the Phishers? Phishing Scam Detection on Ethereum via Network Embedding}}.
\newblock \bibinfo{journal}{{\em {IEEE} Trans. Syst. Man Cybern. Syst.\/}}
  \bibinfo{volume}{52}, \bibinfo{number}{2} (\bibinfo{year}{2022}),
  \bibinfo{pages}{1156--1166}.
\newblock


\bibitem[\protect\citeauthoryear{Xia, Liu, and Wu}{Xia et~al\mbox{.}}{2022}]%
        {XiaYLJW22}
\bibfield{author}{\bibinfo{person}{Yijun Xia}, \bibinfo{person}{Jieli Liu},
  {and} \bibinfo{person}{Jiajing Wu}.} \bibinfo{year}{2022}\natexlab{}.
\newblock
  \showarticletitle{\href{https://doi.org/10.1109/TCSII.2022.3159594}{Phishing
  Detection on Ethereum via Attributed Ego-Graph Embedding}}.
\newblock \bibinfo{journal}{{\em {IEEE} Trans. Circuits Syst. {II} Express
  Briefs\/}} \bibinfo{volume}{69}, \bibinfo{number}{5} (\bibinfo{year}{2022}),
  \bibinfo{pages}{2538--2542}.
\newblock


\bibitem[\protect\citeauthoryear{Xu, Zhang, Vural, Qian, Wang, Fan, Li, Tang,
  and Cao}{Xu et~al\mbox{.}}{2023}]%
        {XuZVQWFLTC23}
\bibfield{author}{\bibinfo{person}{Yufeng Xu}, \bibinfo{person}{Lun Zhang},
  \bibinfo{person}{Turan Vural}, \bibinfo{person}{Peng Qian},
  \bibinfo{person}{Yanbin Wang}, \bibinfo{person}{Yuqing Fan},
  \bibinfo{person}{Ming Li}, \bibinfo{person}{Xueyan Tang}, {and}
  \bibinfo{person}{Zheng Cao}.} \bibinfo{year}{2023}\natexlab{}.
\newblock
  \showarticletitle{\href{https://doi.org/10.1145/3650400.3650499}{STFN:
  Spatio-Temporal Fusion Network to Detect Ethereum Phishing Scams}}. In
  \bibinfo{booktitle}{{\em Proceedings of the 2023 7th International Conference
  on Electronic Information Technology and Computer Engineering, {EITCE}}}.
  \bibinfo{publisher}{{ACM}}, \bibinfo{pages}{599--605}.
\newblock


\bibitem[\protect\citeauthoryear{Yao, Zhang, Xu, Chou, Paturi, Sikder, and
  Saltaformaggio}{Yao et~al\mbox{.}}{2024}]%
        {YaoZHCVAS2024}
\bibfield{author}{\bibinfo{person}{Mingxuan Yao}, \bibinfo{person}{Runze
  Zhang}, \bibinfo{person}{Haichuan Xu}, \bibinfo{person}{Shih{-}Huan Chou},
  \bibinfo{person}{Varun~Chowdhary Paturi}, \bibinfo{person}{Amit~Kumar
  Sikder}, {and} \bibinfo{person}{Brendan Saltaformaggio}.}
  \bibinfo{year}{2024}\natexlab{}.
\newblock
  \showarticletitle{\href{https://doi.org/10.1109/SP54263.2024.00228}{Pulling
  Off The Mask: Forensic Analysis of the Deceptive Creator Wallets Behind Smart
  Contract Fraud}}. In \bibinfo{booktitle}{{\em {IEEE} Symposium on Security
  and Privacy, {SP}}}. \bibinfo{publisher}{{IEEE}},
  \bibinfo{pages}{2236--2254}.
\newblock


\bibitem[\protect\citeauthoryear{Yazdinejad, Dehghantanha, Parizi, Hammoudeh,
  Karimipour, and Srivastava}{Yazdinejad et~al\mbox{.}}{2022}]%
        {YazdinejadDPHKS22}
\bibfield{author}{\bibinfo{person}{Abbas Yazdinejad}, \bibinfo{person}{Ali
  Dehghantanha}, \bibinfo{person}{Reza~M. Parizi}, \bibinfo{person}{Mohammad
  Hammoudeh}, \bibinfo{person}{Hadis Karimipour}, {and} \bibinfo{person}{Gautam
  Srivastava}.} \bibinfo{year}{2022}\natexlab{}.
\newblock
  \showarticletitle{\href{https://doi.org/10.1109/TII.2022.3168011}{Block
  Hunter: Federated Learning for Cyber Threat Hunting in Blockchain-Based IIoT
  Networks}}.
\newblock \bibinfo{journal}{{\em {IEEE} Trans. Ind. Informatics\/}}
  \bibinfo{volume}{18}, \bibinfo{number}{11} (\bibinfo{year}{2022}),
  \bibinfo{pages}{8356--8366}.
\newblock


\bibitem[\protect\citeauthoryear{Yuan, Huang, Zhang, Wu, Zhang, and Zhang}{Yuan
  et~al\mbox{.}}{2020}]%
        {YuanHZWZZ20}
\bibfield{author}{\bibinfo{person}{Qi Yuan}, \bibinfo{person}{Baoying Huang},
  \bibinfo{person}{Jie Zhang}, \bibinfo{person}{Jiajing Wu},
  \bibinfo{person}{Haonan Zhang}, {and} \bibinfo{person}{Xi Zhang}.}
  \bibinfo{year}{2020}\natexlab{}.
\newblock
  \showarticletitle{\href{https://doi.org/10.1109/ISCAS45731.2020.9180815}{Detecting
  Phishing Scams on Ethereum Based on Transaction Records}}. In
  \bibinfo{booktitle}{{\em {IEEE} International Symposium on Circuits and
  Systems, {ISCAS}}}. \bibinfo{publisher}{{IEEE}}, \bibinfo{pages}{1--5}.
\newblock


\bibitem[\protect\citeauthoryear{Zhou, Yang, and Tian}{Zhou
  et~al\mbox{.}}{2023}]%
        {Zhou2023}
\bibfield{author}{\bibinfo{person}{Xuanchen Zhou}, \bibinfo{person}{Wenzhong
  Yang}, {and} \bibinfo{person}{Xiaodan Tian}.}
  \bibinfo{year}{2023}\natexlab{}.
\newblock
  \showarticletitle{\href{https://www.mdpi.com/2079-9292/12/4/993}{Detecting
  Phishing Accounts on Ethereum Based on Transaction Records and EGAT}}.
\newblock \bibinfo{journal}{{\em Electronics\/}} \bibinfo{volume}{12},
  \bibinfo{number}{4} (\bibinfo{year}{2023}).
\newblock


\end{thebibliography}
\end{document}